\documentclass[twocolumn]{article}

\usepackage[utf8]{inputenc}
\usepackage{soulutf8}
\usepackage{subcaption}
\usepackage{booktabs}
\usepackage[separate-uncertainty=true,parse-numbers=true,binary-units=true]{siunitx}
\usepackage{tikz}
\usepackage{pgfplots}
\usepackage{pgfplotstable}
\usepackage{adjustbox}
\usepackage{amsmath}
\usepackage{amssymb}
\usepackage{bm}
\pgfplotsset{compat=1.17}
\usepackage{calc}
\usepackage{nicefrac}
\usepackage{pdfpages}
\usepackage{mathtools}
\usepackage{authblk}

\usepackage{algorithm}
\usepackage{algpseudocode}

\usepackage{geometry}
\newcommand{\extmat}{ in the supplementary material}
\newcommand{\extfig}{ in the supplementary material}

\usepackage{hyperref}
\usepackage[capitalise,nameinlink,noabbrev]{cleveref} 

\usetikzlibrary{quotes,angles,calc,arrows,patterns,quotes}
\newcommand{\diag}{{\rm diag}}

\title{Bayesian Target-Vector Optimization for Efficient Parameter Reconstruction}
\author[a]{Matthias Plock}
\author[b]{Kas Andrle}
\author[a,c]{Sven Burger}
\author[a,c]{Philipp-Immanuel Schneider}

\affil[a]{Zuse Institute Berlin, Takustraße 7, 14195 Berlin, Germany}
\affil[b]{Physikalisch-Technische Bundesanstalt (PTB), Abbestr. 2-12, 10587 Berlin, Germany}
\affil[c]{JCMwave GmbH, Bolivarallee 22, 14050 Berlin, Germany}

\date{}

\newcommand{\mat}{\mathbf}

\renewcommand{\vec}{\boldsymbol}
\newcommand{\vecc}{\boldsymbol}

\newcommand{\trans}{^{\mathrm{T}}}

\begin{document}

\maketitle


\abstract{

  Parameter reconstructions are indispensable in metrology. Here, the objective
  is to to explain $K$ experimental measurements by fitting to them a
  parameterized model of the measurement process. The model parameters are
  regularly determined by least-square methods, i.e., by minimizing the sum of
  the squared residuals between the $K$ model predictions and the $K$
  experimental observations, $\chi^2$. The model functions often involve
  computationally demanding numerical simulations. Bayesian optimization methods
  are specifically suited for minimizing expensive model functions. However, in
  contrast to least-square methods such as the Levenberg-Marquardt algorithm,
  they only take the value of $\chi^2$ into account, and neglect the $K$
  individual model outputs. We present a Bayesian target-vector optimization
  scheme with improved performance over previous developments, that considers
  all $K$ contributions of the model function and that is specifically suited
  for parameter reconstruction problems which are often based on hundreds of
  observations. Its performance is compared to established methods for an
  optical metrology reconstruction problem and two synthetic least-squares
  problems. The proposed method outperforms established optimization methods. It
  also enables to determine accurate uncertainty estimates with very few
  observations of the actual model function by using Markov chain Monte Carlo
  sampling on a trained surrogate model.

}

\section{Introduction}

A common task in science and engineering is the fitting of the parameters of a
model function, in order to match the outputs of the model to an experimental
observation. Often, one has to match $K$ experimental observations
simultaneously, which can be done by minimizing the sum of the squared residuals
$\chi^2$ between model outputs and experimental observations. Generally this is
a non-linear least-squares problem, and is regularly solved by iterative
numerical schemes, e.g., by the Gauss-Newton method or by the Levenberg-Marquardt
algorithm~\cite{leve:1944,marquardt1963algorithm,deuflhard2005newton}.

The models involved in these inverse problems, e.g., in optical
metrology~\cite{vogel2007technology, orji2018natelectron}, are often very costly
to evaluate, because they frequently revolve around the solution of differential
equations via numerical simulations. In scatterometry for example one solves
Maxwell's equations to simulate the scattering process of light off a
nanostructured sample~\cite{raymond2004comparison}. This means that depending on
the complexity and size of the numerical model, a single evaluation can take
several minutes or more to complete. This can present a problem, as many
optimization schemes do not handle expensive model functions in an efficient
way. Typically they calculate the parameters of the next iteration based on only
a few previous model evaluations, neglecting the information from other
evaluations of the optimization history. An additional issue for many model
functions is the difficulty of obtaining their derivatives with respect to the
model parameters. Optimization schemes such as the Levenberg-Marquardt algorithm
require derivatives in order to generate a sampling candidate for the next
iteration. Frequently, derivatives are generated by means of a finite
differences scheme, which is inefficient and often inaccurate. Therefore,
parameter reconstructions in metrology applications can place high demands on
computing resources and
time.
This creates the need for efficient and robust reconstruction methods that make
optimal use of energy and time consuming model evaluations.

Bayesian optimization (BO)~\cite{movckus1975bayesian,mockus2012bayesian} methods
are sequential optimization methods that satisfy the resource requirement, and
are therefore an appropriate choice when optimizing
expensive black-box model functions. At every BO iteration a stochastic model --
most often a Gaussian process (GP)~\cite{williams2006gaussian} -- is trained
using all previous observations of the model function. This stochastic model is
then used to find input parameters for the next iteration, which, e.g., leads to
a large expected improvement over the currently known
minimum~\cite{jones1998efficient}.

Recent studies~\cite{matsui2019bayesian,uhre:2019} considered the extension of
the BO approach for iteratively solving least-squares problems by training $K$
GPs related to each of the $K$ data channels of the model function. In these
works, the approach was successfully applied for problems with $K\leq 20$
channels. Huang et al. \cite{huang2021bayesian} applied these approaches to
calibrate the parameters $\vec{p}\in \mathbb{R}^N$ of an expensive uni-variate
scalar function $f_{\vec{p}}(t)$ to measurements for thousands of times $t$. In
order to reduce the dimensionality of the measurements and the corresponding
functional response to only $K=3$, a functional principal component analysis was
applied, based on $10 N$ space-filling parameter samples prior to solving the
least-square problem. In another study \cite{barnes2020contrasting} a
multi-output GP \cite{alvarez2011kernels,liu2018remarks} was used in a
non-iterative parameter reconstruction approach. I.e., the GP was trained with a
precomputed set of simulation results to directly infer parameter values and
uncertainties. This led to higher reconstruction accuracies than a library
lookup method.

In this study we start from the approach taken by Uhrenholt and
Jensen~\cite{uhre:2019} to develop an algorithm for iterative parameter
reconstruction problems, which often have a large number of data channels $K$,
e.g., a few
hundred~\cite{Hamm:17,szwedowski2019laboratory,kumar2014reconstruction}. We find
that this creates two particular challenges. First, BO methods are typically
associated with a computational overhead of a few seconds when determining new
parameter candidates to sample, which is caused by the training of the surrogate
model. Training and evaluating $K\sim\num{100}$ independent Gaussian processes
dramatically increases this overhead. We propose to use a shared covariance
structure to largely limit the additional overhead. In doing this we assume
implicitly that each data channel has a comparable or similar structure. This
approach will likely lead to deteriorating modeling performance if different
data channel require vastly different length scales to be modeled accurately.
Second, in order to be fast, the scheme approximates the probability
distribution of $\chi^2$ as an ordinary non-central chi-squared distribution
with $K$ degrees of freedom (DoFs). This appears to work well for a small number
of data channels, as is evident from the results presented in~\cite{uhre:2019}.
However, for a large number of DoFs $K$ we find that this leads to a very
inefficient optimization scheme, as the probability distribution largely
underestimates the probability of finding small $\chi^2$ values. We find that
using an \emph{effective} number of DoFs $\tilde K$, which is often much smaller
than $K$, leads to good optimization performance. We propose to choose the value
of $\tilde K$ with maximum-likelihood of the approximate chi-squared probability
distribution for all previous $M$ observations of the model function.

Compared with other approaches, we show that the proposed Bayesian target-vector 
optimization (BTVO) scheme often requires significantly fewer iterations to 
reconstruct the desired model parameters. 
Another important advantage of the approach is that each GP offers a
good non-linear model of the corresponding data channel. This enables, e.g., to
quickly sample from the approximated posterior probability distribution of the
model parameters and to accurately quantify parameter uncertainties.

The paper is organized as follows. In \cref{sec:theory} we give a theoretical
introduction into parameter reconstructions and into solving the corresponding
least-squares problems. We review BO methods and their extension by Uhrenholt
and Jensen. We then discuss the shortcomings for a large number of data channels
$K$ and introduce a mitigation strategy. We then discuss Markov chain Monte
Carlo (MCMC) sampling for determining the uncertainties of the reconstructed
parameters, and we describe a way to drastically reduce the required number of
model evaluations by training and evaluating surrogate models. In
\cref{sec:benchmarks} we compare the performance of the proposed BTVO against a
selection of established optimization methods. We consider three different model
functions: a computationally expensive real-world optical metrology example, and
two analytical model functions obtained from the NIST Standard Reference
Database~\cite{NIST_StRD}. The two analytical model functions are additionally
used to highlight the benefit of derivative information for the reconstruction
performance. In \cref{subsec:mcmc_results} we employ one of the analytical model
functions further to demonstrate the efficiency and accuracy of the surrogate
model augmented MCMC method. Finally, we use the surrogate augmented MCMC to
discuss correlations in the experimental problem discussed in \cite{C8NR00328A}.

\section{Theoretical background}
\label{sec:theory}

Parameter reconstructions are often based on fitting the vectorial output of a
parameterized model function $\vec{f}(\vec{p})$ to an experimental measurement
$\vec{t} = (t_1, \dots, t_K)\trans$, where $\vec{p} \in \mathcal{X} \subset
\mathbb{R}^{N}$ and $\vec{f}: \mathcal{X} \to \mathbb{R}^{K}$. The model
function is treated as a black-box that can be evaluated point-wise, for which
implicitly assume that it is once differentiable and that the $K$ individual
channels can be modeled by Gaussian processes. As usual, we further assume that
the model describes the measurement process sufficiently well, such that model
errors can be neglected. Measurement noise is modeled by assuming that the
$i$-th measurement value is equal to the model value for the true parameter
$\vec{p}_t$ plus some noise contribution,
\begin{equation*}
  \label{eq:meausrement}
  t_i = f_i(\vec{p}_t) + \varepsilon_i \,.
\end{equation*}
Usually, the noise is modeled to be normally distributed with zero
mean and variance $\eta_i^2$, i.e., $\varepsilon_i\sim\mathcal{N}(0,\eta_i^2)$.

Finding a good estimate for $\vec{p}_{t}$ can be considered an optimization
task. We can obtain the least-square estimate (LSQE)
\begin{equation*}
  \vec{p}_{\rm LSQE} =  \underset{\vec{p} \in \mathcal{X}}{\rm arg\,min} \, \chi^2(\vec{p})
\end{equation*}
by minimizing the sum of the squared residuals 
\begin{equation}
  \label{eq:chi_sq}
  \chi^2(\vec{p}) = \sum_{i=1}^{K} \frac{(f_i(\vec{p}) - t_{i})^2}{\eta_i^2}\,.
\end{equation}

Having determined $\vec{p}_{\rm LSQE}$, one is often also interested in the local
probability distribution of the parameter values in order to determine
confidence intervals of the reconstructed parameter values (c.f.
\cref{suppsec:bayesian_param_recon}\extmat).

In optical metrology applications, evaluating the model function $\vec{f}$
typically involves running a numerical simulation, which can include the
assembly of the discretized problem, numerical solution of differential
equations, and postprocessing of the results. Calculating the result of the
model function for a single set of parameters can therefore take a lot of
computation time, depending on the complexity of the model and on numerical
accuracy requirements.

\subsection{Established approaches for parameter reconstruction}
\label{sec:established_approaches}

Minimizing \cref{eq:chi_sq} can be done, e.g., using local methods such as
Nelder-Mead or L-BFGS-B, or using global heuristic methods, for example particle
swarm optimization~\cite{shokooh2007particle,mirjalili2013optical} or
differential evolution~\cite{bor2016differential,saber2017performance}, or by
maximizing the appropriate likelihood function using MCMC sampling
methods~\cite{Herrero:21}. Since minimizing \cref{eq:chi_sq} is a least-square
problem, it can of course also be solved using least-square methods like the
Gauss-Newton scheme or the Levenberg-Marquardt algorithm~\cite{Hamm:17}, which
directly minimize the residuals between model output and experimental results.

An estimate for the confidence intervals can be obtained, e.g., by exploiting
information about the derivatives of the model function with respect to each
parameter at the point estimate $\vec{p}_{\rm LSQE}$, as is for example
available after minimizing \cref{eq:chi_sq} using least-squares
methods~\cite{kutner2005applied,press2007numerical,strutz2010data}, or by
applying MCMC sampling methods to the appropriate likelihood
function~\cite{pfluger2020extracting}. The latter has the advantage that one can
obtain accurate uncertainties in terms of \SI{16}{\percent}, \SI{50}{\percent}
(i.e., the median), and \SI{84}{\percent} percentiles of the actual model
parameter distributions, as well as determine non-linear correlations between
model parameters. Methods that exploit local derivative information usually only
yield approximate Gaussian parameter uncertainties in terms of $1\sigma$
intervals, and are only capable of establishing linear correlations between
model parameters. We denote these Gaussian parameter uncertainties as
$\vecc{\epsilon}_{\rm LSQE}$.

Using these established approaches to perform parameter reconstructions can be a
computationally costly endeavor. Standard least-squares algorithms reconstruct
the first minimum that they reach, which depends on the initial guess fed into
the method. In more complex energy landscapes this can be a local minimum rather
than the point estimate $\vec{p}_{\rm LSQE}$ one seeks, a problem which can be
alleviated, e.g., by using a multi-start approach. This does however not
guarantee that $\vec{p}_{\rm LSQE}$ is found. Particle swarm optimization and
differential evolution on the other hand are not designed to be efficient in the
sense that they obtain the point estimate in as few evaluations of the model
function as possible, and MCMC sampling methods even \emph{rely} on the fact
that they evaluate the model function very often, since the quality of the
reconstruction increases with the number of samples
drawn~\cite{andrieu2003introduction}. This can present an issue from a resource
standpoint, and necessitates more resource efficient optimization methods.

\subsection{Bayesian optimization approaches for parameter reconstruction}
\label{sec:bo}

Bayesian optimization (BO) methods~\cite{movckus1975bayesian,mockus2012bayesian}
are sequential optimization methods and are known for being very efficient at
performing global optimizations of expensive black-box
functions~\cite{Schn:2019Benchmark,jones1998efficient}. To this end BO methods
train a stochastic \emph{surrogate model} -- most often a Gaussian process
(GP)~\cite{williams2006gaussian} -- in an iterative fashion, using all previous
observations of the model function. This surrogate model is usually much quicker
to evaluate than the model function itself. The predictions made by the
surrogate model are then used by an acquisition function to determine a
parameter $\vec{p}_{m+1}$ which is beneficial to sample the model function with
next. The exact meaning of "beneficial" in this context depends very much on the
strategy pursued by the acquisition function, as well as the optimization goal.

In Subsection~\ref{subsec:gpr} we shortly introduce GPs and Gaussian process
regression and describe in Subsection~\ref{subsec:lsq_conv_bo} how it is used in
``conventional'' BO for minimizing $\chi^2(\vec{p})$. In
Subsection~\ref{subsect:lsq_tvo} we will discuss the BTVO approach proposed by
Uhrenholt and Jensen~\cite{uhre:2019}, which minimizes \cref{eq:chi_sq} by
considering the individual contributions of the components of
$\vec{f}(\vec{p})$.

\subsubsection{Gaussian process regression}
\label{subsec:gpr}

GPs are stochastic processes that are defined on a continuous domain
$\mathcal{X} \subset \mathbb{R}^{d}$. A random function $\vec{f}$ is a GP, if
for any finite tuple $\vec{X} = [\vec{x}_{1}, \dots, \vec{x}_{k}] \in
\mathcal{X}^{k}$ the random vector $\vec{Y} = [\vec{f}(\vec{x}_{1}), \dots,
\vec{f}(\vec{x}_{k})]\trans$ is a multivariate normal distribution
(MVN)~\cite{williams2006gaussian}. As such, GPs are an extension of finite
dimensional MVNs to an infinite dimensional case
\cite{garnett_bayesoptbook_2022}. A GP is completely specified by a mean
function $\mu: \mathcal{X} \to \mathbb{R}$ (which replaces a mean vector in the
finite dimensional case) and a covariance kernel function $k: \mathcal{X} \times
\mathcal{X} \to \mathbb{R}$ (replacing a covariance matrix of a finite
dimensional MVN)~\cite{williams2006gaussian}. A GP can be trained to determine
the posterior distribution of function values given some observations of the
function, and can then serve as a stochastic predictor or interpolator for the
training data. A usual choice for the prior mean and covariance kernel function,
which is also considered in the following, are a constant mean function and the
Matérn $5/2$ covariance function~\cite{brochu2010tutorial}
\begin{gather}
  \begin{aligned}[t]
    \mu(\vec{p}) &= \mu_0 \,, \nonumber \\
    k(\vec{p},\vec{p}') &=
    \sigma_0^2\left(1+\sqrt{5}r+\frac{5}{3}r^2\right)\exp\left(-\sqrt{5}r\right)
    \,, \nonumber 
  \end{aligned}\\
  \text{where} \quad r = \sqrt{\sum_{i=1}^N
    \frac{(p_i-p_i')^2}{l_i^2}} \label{eq:r}\,.
\end{gather}
Up to some maximum number of observations $M_{\rm hyper}$ the hyperparameters
$\mu_0, \sigma_0, l_1, \dots, l_N$ are chosen to maximize the likelihood of the
observations~\cite{sant:2018,sant:2021}. Afterwards, only $\mu_0, \sigma_0$ are
optimized and the length scales $l_1, \dots, l_N$, which enter the covariance 
function in a non-trivial way, are kept constant.
A GP trained on function evaluations
$\vec{Y} = \left[f(\vec{p}_1), \dots, f(\vec{p}_M)\right]\trans$
allows to make predictions for any parameter vector $\vec{p}^\ast$ in the form
of a normally distributed random variable $\hat{f}(\vec{p}^\ast) \sim
\mathcal{N}(\overline{y}(\vec{p}^\ast),\sigma^2(\vec{p}^\ast))$ with mean and
variance
\begin{align}
  \label{eq:pred_mean}
  \overline{y}(\vec{p}^\ast) &= \mu_0 + \vec{k}\trans (\vec{p}^\ast)
                               \mat{K}^{-1}[\vec{Y}-\mu_0 \vec{1}] \\
  \label{eq:pred_var}
  \sigma^2(\vec{p}^\ast) &= \sigma_0^2 - \vec{k}\trans (\vec{p}^\ast)
                           \mat{K}^{-1} \vec{k}(\vec{p}^\ast) \,,
\end{align}
where $\vec{k}(\vec{p}^\ast) =
\left[k(\vec{p}^\ast,\vec{p}_1),\dots,k(\vec{p}^\ast,\vec{p}_M)\right]^{\rm T}$
and $(\mat{K})_{i j} = k(\vec{p}_i,\vec{p}_j)$.

For better numerical stability, the positive semidefinite covariance matrix
$\mat{K}$ is not inverted directly. Instead, one can compute its Cholesky
decomposition $\mat{K} = \mat{L}_{\mat{K}} \mat{L}_{\mat{K}}\trans$ into a lower
and upper triangular matrix in $\mathcal{O}(M^3)$ steps for $M\leq M_{\rm
  hyper}$. For constant length scales (i.e., $M > M_{\rm hyper}$) an update of
the decomposition only requires $\mathcal{O}(M^2)$ steps~\cite{sant:2021}.
Afterwards, one solves
\begin{equation}
  \label{eq:alpha}
  \mat{L}_{\mat{K}} \mat{L}_{\mat{K}}\trans \vecc{\alpha} = \vec{Y}-\mu_0 \vec{1}
\end{equation} 
for $\vecc{\alpha}$ by forward and backward substitution and
\begin{equation}
  \label{eq:beta}
  \mat{L}_{\mat{K}}\vecc{\beta}(\vec{p}^\ast) = \vec{k}(\vec{p}^\ast)
\end{equation}
for $\vecc{\beta}(\vec{p}^\ast)$ by forward substitution both in
$\mathcal{O}(M^2)$ steps. With the auxiliary vectors $\vecc{\alpha}$ and
$\vecc{\beta}$, \cref{eq:pred_mean,eq:pred_var} can be evaluated as
$\overline{y}(\vec{p}^\ast) = \mu_0 + \vec{k}(\vec{p}^\ast)\trans\vecc{\alpha}$
and $\sigma^2(\vec{p}^\ast) = \sigma_0^2 - \vecc{\beta}\trans\vecc{\beta}$ in
only $\mathcal{O}(M)$ steps. For the computation of the next sampling point, one
requires predictions for many points $\vec{p}^\ast$. Hence, solving
\cref{eq:beta} for $A$ different values of $\vec{p}^\ast$ (typically $A\gtrsim
1000$) requires a large fraction of the computation time.

Training of GPs can be easily extended to exploit derivative information if
available. When solving Maxwells equations this can be calculated e.g. using
  the direct method or the adjoint method~\cite{sant:2021}. While this enlarges
the number of data points and the size of the covariance matrix accordingly, all
the above considerations are equally valid.

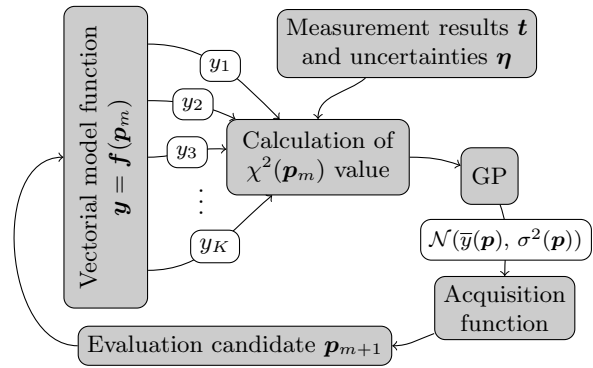
\begin{figure}[t]
  \centering




\begin{tikzpicture}[font=\small, scale=1]

  \node[draw, rectangle, minimum height=1.1cm, minimum width=4cm, fill=black!20,
  rounded corners, rotate=90, align=center] at (-0.25, 0) (sim) {Vectorial model
    function \\$\vec{y} = \vec{f}(\vec{p}_{m})$};
  
  
  \node[draw, fill=black!20, rounded corners, align=center] at
  ([xshift=2.25cm,yshift=-0.5cm]sim.east)[right] (results) {Measurement results
    $\vec{t}$ \\ and uncertainties $\vecc{\eta}$};

  \node[draw, fill=black!20, rounded corners, align=center, rectangle, minimum
  height=1cm, minimum width=2.4cm] at (3.75, 0.0)[left] (chisq) {Calculation of
    \\ $\chi^{2}(\vec{p}_{m})$ value};

  \node[draw, fill=black!20, rounded corners, align=center, rectangle, minimum height=.75cm, minimum width=.75cm] at ([xshift=1.05cm,yshift=-0.25cm]chisq.east) (gp)
  {GP};

  \foreach \i in {1,...,3}{%
        
    \path[->] ([yshift=2.25cm - \i * 0.75 cm]sim.south) edge[out=0,in=115 + \i * 20]
    node[font=\footnotesize, draw, rounded corners, pos=.5, fill=white, opacity=1, text opacity=1] (p\i) {$y_{\i}$} (chisq);

    ;}
  
  \path[->] ([yshift=-1.5 cm]sim.south) edge[out=0,in=220] node[draw, rounded corners, pos=.5, fill=white, opacity=1, text opacity=1] (pN)
  {$y_{K}$} (chisq);

  \node[rotate=90] at ($(p3)!0.5!(pN)$) (dots) {\dots};

  \node[draw, fill=black!20, rounded corners, align=center] at (5, -2)
  (af) {Acquisition \\ function};


  \path[->] (results) edge[out=220,in=90] (chisq);

  \path[->] (chisq.east) edge[out=0,in=160] (gp);

  \path[->] (gp) edge[out=290,in=90] node[font=\footnotesize, draw, rounded corners, pos=.5, fill=white, opacity=1, text opacity=1] {$\mathcal{N}(\overline{y}(\vec{p})$, $\sigma^2(\vec{p}))$} (af);

  \node[draw, fill=black!20, rounded corners, align=center] at (1.45, -2.5)
  (cand) {Evaluation candidate $\vec{p}_{m+1}$};

  \path[->] (af) edge[out=200, in=0] (cand);

  \path[->] (cand.west) edge[out=180,in=180] (sim.north);
  
\end{tikzpicture}
  \caption{Schematic of a least-squares fit using the conventional Bayesian
    optimization method. The multiple outputs of the actual model are used to
    calculate the value of $\chi^2(\vec{p}_{m})$ in \cref{eq:chi_sq} for some
    parameter $\vec{p}_{m}$, which is then used to train the surrogate model (in
    this case a Gaussian process, GP). This surrogate predicts a normal distribution
    for each point in the parameter space, which is used by the acquisition
    function. The acquisition function determines a candidate parameter
    $\vec{p}_{m+1}$ which is used to evaluate the actual model function again.
  }
  \label{fig:schematic_bo}
\end{figure}

\subsubsection{Bayesian optimization}
\label{subsec:lsq_conv_bo}

First, we consider the conventional BO method in order to minimize the scalar
function $\chi^2(\vec{p})$ defined in \cref{eq:chi_sq}. At each iteration $m$,
the BO approach employs the predictions made by the trained GP in order to
determine the next sampling point $\vec{p}_{m+1}$. This point is selected
according to some infill criterion at the maximum of an acquisition function
$\alpha(\vec{p})$~\cite{movckus1975bayesian,mockus2012bayesian}. A usual choice
is the expected improvement (EI) with respect to the lowest known function value
$\chi^2_{\rm min} = \min \{\chi^2(\vec{p}_1), \dots, \chi^2(\vec{p}_M)\}$.

The corresponding acquisition function is defined as
\begin{equation}
  \label{eq:ei}
  \alpha_{\rm EI}(\vec{p}) = \mathbb{E}\left[\min(0, \chi^2_{\rm min} - \hat{f}(\vec{p}))\right],
\end{equation} 
where $\hat{f}(\vec{p})$ is a Gaussian random variable with mean
$\overline{y}(\vec{p})$ and variance $\sigma^2(\vec{p})$ given in
\cref{eq:pred_mean,eq:pred_var}. Another infill criterion is the lower
confidence bound (LCB) with acquisition function
\begin{equation}
  \label{eq:lcb}
  \alpha_{\rm LCB}(\vec{p}) =  \kappa \sigma^2(\vec{p}) - \overline{y}(\vec{p}),    
\end{equation}
where $\kappa$ is a scaling factor.

\subsubsection{Bayesian target-vector optimization}
\label{subsect:lsq_tvo}

Directly minimizing the squared error function \cref{eq:chi_sq} using BO (c.f.
\cref{fig:schematic_bo}) has some important drawbacks. First, it ignores the
knowledge of the function values $f_1(\vec{p}),\dots,f_K(\vec{p})$ that
contribute to the value of $\chi^2(\vec{p})$ in \cref{eq:chi_sq}. This
information loss leads to a significantly slower convergence. The second issue
stems from the fact that a GP can only be used to predict normal distributions,
i.e., \cref{eq:pred_mean,eq:pred_var}. The predictions of a GP that has been
conditioned on observations of \cref{eq:chi_sq} are necessarily incorrect, since
$\chi^2(\vec{p})$ does not follow a normal distribution, but rather a
chi-squared distribution. This problem becomes especially pronounced when the GP
predicts small mean values and large variances, as this may lead the acquisition
function into exploring regions where the predictions suggest an improvement to
negative $\chi^2$ mean values. This clearly conflicts with $\chi^2$ being larger
than, or equal to, zero. To address the issues, we follow the approach proposed
by Uhrenholt and Jensen~\cite{uhre:2019}, in which each of the $K$ components of
$\vec{f}(\vec{p})$ is modeled by a GP.

Using $K$ independent GPs for making a regression on $K$ channels increases the
computational effort to $\mathcal{O}(K\cdot M^3)$ steps for computing $K$
Cholesky decompositions. For $M > M_{\rm hyper}$ observations we keep the length
scales constant and the computational effort increases to $\mathcal{O}(A\cdot
K\cdot M^2)$ steps for making $A$ different prediction in order to maximize the
acquisition function. For metrology applications with often more than 100
channels, the corresponding computation times render the approach impractical.
Therefore, in contrast to the approach of Uhrenholt and Jensen, we propose to
model the GPs using the same covariance kernel function and only allow for
different hyperparameters $\mu_0^{(i)}$ and $\sigma_0^{(i)}$ for each channel
$i=1,\dots,K$. The optimal value of the hyperparameters $\mu_0^{(i)}$,
$\sigma_0^{(i)}$ for $i = 1, \dots, K$ and the length scale parameters $l_1,
\dots, l_N$ for the shared covariance kernel matrix are chosen by maximizing the
likelihood of the training data averaged over all channels. Hence, the Cholesky
decomposition has to be computed only once and can be used for all channels as
well as for the solution of \cref{eq:beta}. Only \cref{eq:alpha} has to be
solved for all $K$ channels since it depends on the acquired function values
$\vec{Y}$ in each channel. However, the equation has to be solved only once to
make an arbitrary number $A$ of predictions for different parameter vectors
$\vec{p}^\ast$. Since $A$ is usually much larger than $K$, the computational
overhead of $\mathcal{O}(K\cdot M^2)$ steps for solving \cref{eq:alpha} for each
channel turns out to being acceptable in comparison to solving \cref{eq:beta}
repeatedly in $\mathcal{O}(A\cdot M^2)$ steps.

The $K$ GPs each provide predictions in form of a normally distributed random
variable
$\hat{f}_i(\vec{p})\sim\mathcal{N}(\overline{y}_i(\vec{p}),\sigma^2_i(\vec{p}))$,
for $i = 1,\dots,K$, such that the prediction of $\chi^2(\vec{p})$ defined in
\cref{eq:chi_sq} is a random variable
\begin{equation}
  \label{eq:mogp_chi_square_prediction}
  \hat\chi^2(\vec{p}) = \sum_{i=1}^{K} \frac{(\hat{f}_i(\vec{p}) - t_i)^2}{\eta_i^2} \,.
\end{equation}
which follows a \emph{generalized} chi-squared distribution. We note, that
modeling $\hat{f}_1(\vec{p}),\dots,\hat{f}_K(\vec{p})$ as independent random
variables neglects correlations between the channels. However, modeling these
correlations by training of a joint GP for all channels is computationally very
expensive since the covariance matrix of the joint GP has $K^2 \cdot M^2$
instead of just $M^2$ entries.

\begin{figure}[t]
  \centering



\begin{tikzpicture}[font=\small]

  \node[draw, rectangle, minimum height=1.1cm, minimum width=4cm, fill=black!20,
  rounded corners, rotate=90, align=center] at (-0.25, 0) (sim) {Vectorial model function \\$\vec{y} = \vec{f}(\vec{p}_{m})$};

  
  \node[draw, fill=black!20, rounded corners, align=center] at
  ([xshift=.75cm,yshift=0.5cm]sim.east)[right] (results) {Measurement results
    $\vec{t}$ \\ and uncertainties $\vecc{\eta}$};

  \foreach \i in {1,...,3}{%
    
    \node[draw, fill=black!20, rounded corners, align=center] at
    ([xshift=1.35cm, yshift=2.25cm - \i * 0.75 cm]sim.south) (gp\i)
    {$\mathrm{GP}_{\i}$};
    
    \path[->] ([yshift=2.25cm - \i * 0.75 cm]sim.south) edge[out=0,in=180]
    node[font=\footnotesize, draw, rounded corners, pos=.5, fill=white, opacity=1, text opacity=1] {$y_{\i}$} (gp\i);

    ;}

  \node[draw, fill=black!20, rounded corners, align=center] at
  ([xshift=1.35cm, yshift=-1.5 cm]sim.south) (gpN)
  {$\mathrm{GP}_{K}$};
  
  \path[->] ([yshift=-1.5 cm]sim.south) edge[out=0,in=180] node[font=\footnotesize, draw, rounded corners, pos=.5, fill=white, opacity=1, text opacity=1]
  {$y_{K}$} (gpN);

  \node[rotate=90] at ($(gp3)!0.5!(gpN)$) (dots) {\dots};
  
  \node[draw, fill=black!20, rounded corners, align=center] at (6.25, 0)[left]
  (pdf) {Calculation of \\ $\hat\chi^2(\vec{p})$ prediction};

  \path[->] (results.east) edge[out=0,in=90] (pdf);

  \path[->] (gp1.east) edge[out=0,in=105 + 25] node[font=\scriptsize, draw, rounded corners, pos=.45, fill=white, opacity=1, text opacity=1] {$\mathcal{N}(\overline{y}_{1}(\vec{p})$, $\sigma^2_{1}(\vec{p}))$} (pdf);

  \foreach \i in {2,...,3}{%
    
    \path[->] (gp\i.east) edge[out=0,in=105 + \i * 25] node[font=\scriptsize, draw, rounded corners, pos=.5, fill=white, opacity=1, text opacity=1] {$\mathcal{N}(\dots)$} (pdf);

    ;}

  \path[->] (gpN.east) edge[out=0,in=230] node[above, font=\scriptsize, draw, rounded corners, pos=.5, fill=white, opacity=1, text opacity=1] {$\mathcal{N}(\overline{y}_{K}(\vec{p})$, $\sigma^2_{K}(\vec{p}))$} (pdf);

  \node[draw, fill=black!20, rounded corners, align=center] at (5, -2)
  (acquisition) {Acquisition \\ function};

  \node[draw, fill=black!20, rounded corners, align=center] at (1.5, -2.5)
  (cand) {Evaluation candidate $\vec{p}_{m+1}$};

  \path[->] (pdf) edge[out=280, in=80] (acquisition);

  \path[->] (acquisition) edge[out=200, in=0] (cand);
  
  \path[->] (cand.west) edge[out=180,in=180] (sim.north);
  
\end{tikzpicture}
  \caption{Schematic of a least-squares fit using the Bayesian target-vector
    optimization method. The $K$ outputs of the actual model function are used
    to train $K$ surrogate models (in this case $K$ Gaussian processes), each of
    which predicts a normal distribution for each point in the parameter space,
    which are used to calculate a predicted $\hat\chi^2$ distribution. This
    predicted distribution is then used by the acquisition function, which in
    turn determines a candidate parameter $\vec{p}_{m+1}$, which is used to
    evaluate the actual model function again.}
  \label{fig:schematic_tvo}
\end{figure}
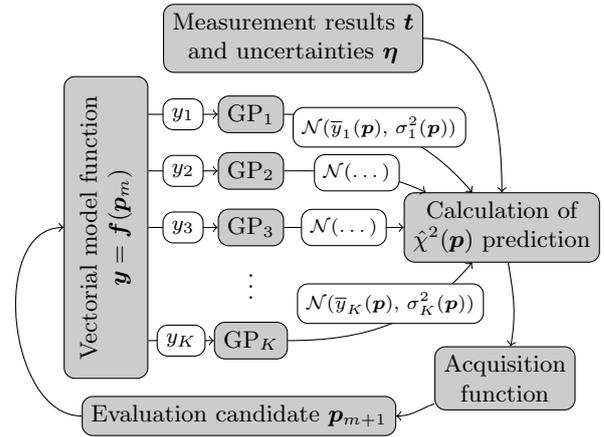

The values of the probability distribution function (PDF) and cumulative
distribution function (CDF) of $\hat\chi^2(\vec{p})$ are in general expensive to
calculate~\cite{mathai1992quadratic,mohsenipour2012distribution}. In order to be
able to efficiently compute the distribution, Uhrenholt and Jensen resort to
approximating the generalized chi-squared probability distribution by a
\emph{non-central} chi-squared distribution of the renormalized random variable
$\hat{\chi}^2_{\rm ren}(\vec{p}) \approx \hat\chi^2(\vec{p})/\gamma^2(\vec{p})$
with $K$ DoFs, average normalized variance
\begin{equation}
  \label{eq:gammasq}
  \gamma^2(\vec{p}) =\frac{1}{K}\sum_{i=1}^{K} \frac{\sigma_i^2(\vec{p})}{\eta_i^2} \,,
\end{equation}
and non-centrality parameter
\begin{equation}
  \label{eq:lambda}
  \lambda(\vec{p}) = \frac{1}{\gamma^2(\vec{p})} \sum_{i=1}^{K} \frac{\left(\overline{y}_i(\vec{p}) - t_i\right)^2}{\eta_i^2} \,.
\end{equation}
This approximation is now followed by a second approximation, in which the
non-central chi-squared distribution is approximated by a parameterized normal
distribution. This makes it possible to evaluate the employed LCB infill
criterion \cref{eq:lcb} (we use a constant scaling factor $\kappa = 3$) of the
multi-output GP prediction \cref{eq:mogp_chi_square_prediction} with good
efficiency and accuracy~\cite{uhre:2019}. In a previous publication we
successfully applied the described method to a parameter reconstruction problem
for a relatively small number of data channels
$K$~\cite{plock_et_al_recent_advances}. \cref{fig:schematic_tvo} shows a
schematic of the BTVO being used to minimize \cref{eq:chi_sq}.

In order to demonstrate the issues that arise for a large number of data
channels for this series of approximations, we consider a Wilson-Hilferty
approximation of the non-central chi-squared distribution, according to which
the random variable
\begin{equation*}
  \hat{z}(\vec{p}) = \left( \frac{\gamma^{-2}(\vec{p})\hat\chi^2(\vec{p})}{K + \lambda(\vec{p})} \right)^{\nicefrac{1}{3}}
\end{equation*}
approximately follows a normal distribution~\cite{abdel:1954} with mean
$\overline{z}(\vec{p}) = 1-\nicefrac{2}{9\nu(\vec{p})}$ and variance
$\sigma^2_{z}(\vec{p}) = \nicefrac{2}{9\nu(\vec{p})}$, where $\nu(\vec{p}) =
\nicefrac{(K+\lambda(\vec{p}))^2}{K+2\lambda(\vec{p})}$. We consider the limit
of $K \to \infty$ while keeping the non-centrality $\lambda(\vec{p})/K$ per
channel constant. This is a good approximation for the case that the density of
data channels in an experiment, e.g., the density of a measured spectrum, is
increased. When increasing $K$, we observe that $\sigma_z^2(\vec{p}) \to 0$,
which means that the PDF approaches a $\delta$-distribution and the CDF a step
function. Since $\overline{z}(\vec{p}) \to 1$, the maximum of the PDF approaches
\begin{multline}
  \label{eq:ren_mode}
  \chi^2_{\rm mode}(\vec{p}) \to \gamma^2(\vec{p}) (K + \lambda(\vec{p})) \\
  = \sum_{i=1}^K \frac{(\overline{y}_i(\vec{p}) - t_i)^2 +
    \sigma^2_i(\vec{p})}{\eta_i^2} \,.
\end{multline}
Let $\vec{p}_{\rm min}$ be the parameter with the minimal observed value of
$\chi^2$. Moving away from $\vec{p}_{\rm min}$ to regions with fewer training
samples, $\overline{y}_i(\vec{p})$ approaches the mean value of the
corresponding GP, $\mu_0^{(i)}$, which has generally a larger deviation from the
measurement $t_i$. Moreover the GP uncertainty increases. Hence, using this
approximated probability function for large $K$ the probability density of
seeing $\chi^2$ values smaller or close to $\chi^2(\vec{p}_{\rm min})$ quickly
approaches zero for points not close to $\vec{p}_{\rm min}$. Consequently, the
employed infill criterion can only select points close to $\vec{p}_{\rm min}$.
In contrast to conventional BO, this leads to a very localized optimization
behavior with possibly slow convergence and no exploration of regions with fewer
training data. The selection of sample candidates that are too close to
previously sampled positions also has a negative impact on the numerical
stability of the scheme, as it leads to an ill-conditioned covariance matrix
$\mat{K}$. Therefore, the optimization is terminated if the infill criterion is
only able to select samples that are within a distance of $\num{1e-3}$ length
scales to previously sampled positions.

This undesired behavior of the approximate probability distribution for large
$K$ is a consequence of modeling all channels by independent random variables.
For parameter reconstruction problems we know, however, that there exist
parameter values for which \emph{all} $K$ residuals are small at the same time
-- an event that is expected to be almost impossible if $K$ is large. Moreover,
channels that belong to similar experimental conditions (e.g., similar angles
and wavelengths in scatterometry experiments) show a similar dependence on the
parameters $\vec{p}$ and are thus positively correlated. If sum rules apply to
the measurement process (e.g., energy or momentum conservation) also negative
correlations between the channels can be observed. As described above,
neglecting these $\vec{p}$-dependent correlations is done for performance
reasons.

To mitigate these issues, we propose to replace the number of DoFs $K$ in the
parameterization of the approximated probability distribution by an
\emph{effective} number of DoFs $\tilde K$, such that the approximate
probability distribution better matches \emph{all} $M$ previous observations
$\vec{Y}$ of the model function. To this end, we take the sum over all
$\chi^2_{m}$ values, $\chi^2_{\rm all} = \sum_{m=1}^{M} \chi^2_m$, find the
corresponding approximate marginal likelihood distribution, and then maximize
the likelihood of the observed $\chi^2_{\rm all}$ value with respect to the
effective DoFs to obtain the maximum-likelihood estimate $K_{\rm MLE}$. The
acquisition function is then evaluated based on a non-central chi-squared
distribution with $\tilde K = K_{\rm MLE}$ effective DoFs and non-centrality
defined in \cref{eq:lambda}. The derivation of $\tilde K$ is detailed in
\cref{suppsec:effdof}\extmat. We find that this yields an effective way to
regain the exploratory nature of the BO method also for the BTVO with a large
number of data channels $K$. A complete algorithmic overview of the BTVO scheme
can be found in \cref{alg:btvo}\extmat.

\subsection{Markov chain Monte Carlo for accurate parameter uncertainties}
\label{sec:mcmc}

By exploiting information from the Jacobian of the model function at the LSQE
$\vec{p}_{\rm LSQE}$ it is possible to give estimates for the parameter
uncertainties $\vec{\epsilon}_{\rm LSQE}$, c.f. also
\cref{suppsec:levenberg-marquardt}\extmat. These uncertainty estimates are
generally symmetrical, which does not necessarily reflect on the true model
parameter distribution. Furthermore, it is only possible to obtain approximate
linear correlations between model parameters, while one is often interested in
more complex model parameter relationships when performing parameter
reconstructions~\cite{pfluger2020extracting}.

More accurate parameter uncertainties beyond the Gaussian approximation can be
given in terms of \SI{16}{\percent}, \SI{50}{\percent}, and \SI{84}{\percent}
percentiles of the likelihood $\mathcal{L}(\vec{p})$ or their posterior
probability $\mathcal{P}(\vec{p})$ of the parameter values
(c.f.~\cref{suppsec:bayesian_param_recon}\extmat). The \SI{50}{\percent}
percentile, i.e. the median, then represents the maximum likelihood estimate
(MLE) (when sampling the likelihood function) or the maximum a-posteriori
estimate (MAP) (when sampling the posterior density).

Unfortunately, for the non-Gaussian probability distributions
$\mathcal{L}(\vec{p})$ and $\mathcal{P}(\vec{p})$, the percentiles can in
general not be determined analytically. Instead, one determines them based on a
large set of samples whose density in the parameter space is proportional to the
probability distribution of interest. These sets are typically generated by
Markov chain Monte Carlo (MCMC) sampling
methods~\cite{andrieu2003introduction,sammut2011encyclopedia,friedman2001elements}.
The sampling sets can be also used to expose non-linear correlations between the
model parameters~\cite{pfluger2020extracting} and allow to reconstruct
non-trivial error models from the observed data~\cite{Henn:12}. This makes MCMC
a valuable tool for parameter reconstructions in general, not just in the field
of optical metrology.

In practice, MCMC often requires tens of thousands of evaluations of
$\mathcal{L}(\vec{p})$ or $\mathcal{P}(\vec{p})$, each associated with an
evaluation of the model function $\vec{f}(\vec{p})$, to construct a stable
equilibrium distribution. For computationally expensive model evaluations, this
can require very large computational resources.

\subsubsection{Markov chain Monte Carlo on a Gaussian process surrogate}
\label{subsec:surrogate_mcmc}

To lessen the resource impact of MCMC sampling, it can be favorable to use a
trained and quick-to-evaluate surrogate model of the actual model function to
construct the equilibrium distribution Markov chain. Recently, MCMC using
surrogate models based on a polynomial chaos
expansion~\cite{farchim_efficient_inversion} as well as invertible neural
networks~\cite{invertible_nn_andrle} were proposed and showed to be converging
to the ``exact'' results, where ``exact'' in this context means that MCMC was
performed on the model function directly.

We propose to instead utilize the multi-output GP surrogate model that was
trained during a parameter reconstruction with the BTVO scheme. It is used to
calculate the predicted likelihood function $\hat{\mathcal{L}}(\vec{p})$ of the
Gaussian prediction $\hat{\vec{f}}(\vec{p})$ of the model function, which is
then used to generate the equilibrium distribution. We have to assert that the
surrogate model is an accurate representation of the model function in the
region of interest, which is often located around the acquired LSQE point.
Because the surrogate model was created during a parameter reconstruction, this
region is often only explored reasonably well in the direction from which the
LSQE was found. To reduce the uncertainty of the surrogate model in the rest of
the region of interest, we enter a refinement stage in which we actively train
the model with more parameter samples close to the LSQE point $\vec{p}_{\rm
  LSQE}$. To this end, we draw $S$ random samples $\vec{p}_1,\dots,\vec{p}_S$
from the multivariate normal distribution $\mathcal{N}(\vec{p}_{\rm
  LSQE},\mat{Cov}(\vec{p}_{\rm LSQE}))$ defined by the parameter covariance
matrix $\mat{Cov}(\vec{p}_{\rm LSQE})$ at the LSQE point
(c.f.~\cref{suppsec:levenberg-marquardt}\extmat) and evaluate the forward model
at the point with maximum mean uncertainty of the surrogate models of all
channels, i.e.,
\begin{equation*}
  \vec{p}_{\rm max} = \underset{\vec{p} \in
    \{\vec{p}_1,\dots,\vec{p}_S\}}{\rm arg\,max}\frac{1}{K}\sum_{k=1}^K
  \sigma_k(\vec{p}) \,.
\end{equation*}
In essence we follow a sequential experimental design strategy in which the
criterion is the predicted variance of the surrogate model
\cite{gramacy2020surrogates}. We stop the additional sampling of the forward
model when the maximum mean uncertainty is below some threshold $\sigma_{\rm
  min}$. An algorithmic overview can be found in \cref{alg:mcmc}\extmat.

\section{Benchmarks}
\label{sec:benchmarks}

In order to assess the performance of the proposed BTVO scheme, we applied it to
three parameter reconstruction problems: an experimental dataset, where the
model function is a finite element simulation, and two synthetic datasets with
analytic model functions. 

The experimental dataset has been measured during a Grazing Incidence X-Ray
Fluorescence (GIXRF) experiment at the synchrotron radiation source BESSY in
Berlin \cite{C8NR00328A}. The analytic datasets were obtained from the NIST
Standard Reference Database~\cite{NIST_StRD} for non-linear regression problems.
To show resilience of the reconstruction algorithm, we chose two datasets that
differ with respect to the number of free parameters and data points: the MGH17
dataset~\cite{NIST_MGH17} contains five free parameters and 33 data points,
while the Gauss3 dataset~\cite{NIST_Gauss3} contains eight free parameters and
250 data points. The Gauss3 dataset is of particular interest, as it is
comparable to problems from optical metrology in terms of free parameters and
data points. Because of their analytical nature, derivatives with respect to all
free parameters are easily calculated.

The proposed BTVO method was compared in a benchmark type analysis, where its
reconstruction performance was compared to that of other optimization schemes.
Here, Levenberg-Marquardt (LM)~\cite{leve:1944,marquardt1963algorithm,flet:1971}
(also c.f. \cref{suppsec:levenberg-marquardt}\extmat, for the benchmarks we
employed the scipy implementation
\verb|scipy.optimize.least_squares|~\cite{2020SciPy-NMeth}), BO as detailed in
\cref{subsec:lsq_conv_bo}, the limited memory Broyden–Fletcher–Goldfarb–Shanno
algorithm with box constraints (L-BFGS-B)~\cite{byrd1995limited}, as well as the
Nelder-Mead (NM) downhill simplex algorithm~\cite{nelder1965simplex} were used.

Of these methods, only BTVO and LM are methods that can solve least-squares
problems natively by utilizing all data channels of the model function. BO,
L-BFGS-B and NM are optimization methods that are usually used for the
minimization or maximization of scalar functions. These methods therefore
minimize (functions of) $\chi^2(\vec{p})$ as defined in \cref{eq:chi_sq}
directly. L-BFGS-B and NM both took the regular scalarized value
$\chi^2(\vec{p})$ as model function, where L-BFGS-B explicitly profits from this
as it works best on functions that are quadratic~\cite{liu1989limited}. The
surrogate model employed by BO assumes that the model function outputs are
normally distributed. By taking the third root of a chi-squared distributed
random variable it can be transformed into a more normally distributed random
variable~\cite{hawkins1986note}. Therefore, BO minimized
$(\chi^2)^{\nicefrac{1}{3}}(\vec{p})$ instead of $\chi^2(\vec{p})$. Of the
investigated methods only the Nelder-Mead algorithm can not take advantage of
derivative information if provided.

For each reconstruction method and dataset, six consecutive optimizations were
performed. From these results we calculated a mean and a standard deviation of
the respective reconstruction result. The metric chosen to quantify the
reconstruction performance of each method was the distance $d(\vec{p})$ of the
reconstructed parameter to the LSQE point $\vec{p}_{\rm LSQE}$, given in units
of the reconstructed Gaussian model parameter uncertainties $\vec{\epsilon}_{\rm
  LSQE}$, i.e.,
\begin{equation}
    \label{eq:distance}
    d(\vec{p}) = \sqrt{ \sum_{i=1}^M \left( \frac{p_i - p_{{\rm LSQE},i}}{\epsilon_{{\rm LSQE},i}} \right)^2 } \,.
\end{equation}
We considered the reconstruction to be converged if parameter values with a
distance $d < \num{0.1}$ were found, i.e., the parameters deviated only by
\SI{10}{\percent} of the confidence interval. Note that the values for
$\vec{p}_{\rm LSQE}$ and $\vec{\epsilon}_{\rm LSQE}$ are determined after all
optimizations are finished. $\vec{p}_{\rm LSQE}$ is determined from the smallest
$\chi^2$ value for each iteration across all optimization runs and schemes in
the benchmark, the value for $\vec{\epsilon}_{\rm LSQE}$ is determined from the
Jacobian for that parameter.

To demonstrate the positive impact of using the effective DoFs
$\tilde K$ to parameterize the predictive distribution in the BTVO scheme, we
performed the same benchmarks as described above, but instead using the number
of data channels $K$ to parameterize the predictive distribution.

Afterwards, we assessed the efficiency of surrogate model augmented MCMC
sampling by applying it to the MGH17 dataset, and compared the results to those
of MCMC sampling applied to the exact likelihood function. Finally, we briefly
discussed the findings of using the surrogate augmented MCMC on the experimental
GIXRF dataset.

The reconstruction of the GIXRF problem was performed on a Dell PowerEdge R7525
Rack Server with 2x AMD EPYC 7542 32-core CPUs (yielding 128 usable threads) and
1 TB of RAM installed. The reconstructions of the two analytical problems were
done on a workstation computer with a AMD Ryzen 7 3700X 8-core CPU and 32 GB of
RAM.

The Bayesian methods employed are implemented in the analysis and optimization
toolkit of the commercial finite element Maxwell solver
JCMsuite~\cite{JCMsuite}. The research data for this manuscript will be released
separately~\cite{btvo_paper_zenodo}.

\subsection{The experimental dataset: Grazing Incidence X-Ray Fluorescence
  (GIXRF)}
\label{subsec:gixrf}

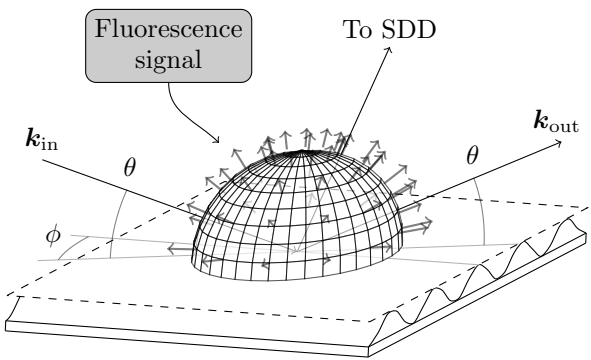
\begin{figure}[ht]
    \centering
    \begin{tikzpicture}[xscale=1.3, yscale=1.3]

  \begin{scope}[
    rotate around y = -10,
    rotate around x = -95
    ]
    \def\h{5}
    \def\pi{3.1415}

    \newcommand{\quadrant}[2]{
      \foreach \t in {#1} \foreach \f in {5,15,...,90}
      \draw [fill=#2,opacity=0.65]
      ({sin(\f - \h)*cos(\t - \h)}, {sin(\f - \h)*sin(\t - \h)}, {cos(\f - \h)})
      -- ({sin(\f - \h)*cos(\t + \h)}, {sin(\f - \h)*sin(\t + \h)}, {cos(\f - \h)})
      -- ({sin(\f + \h)*cos(\t + \h)}, {sin(\f + \h)*sin(\t + \h)}, {cos(\f + \h)})
      -- ({sin(\f + \h)*cos(\t - \h)}, {sin(\f + \h)*sin(\t - \h)}, {cos(\f + \h)})
      -- cycle;
    }

    \draw (-2, -2, -0.35) -- (2, -2, -0.35) -- (2, 2, -0.35);
    \draw (-2, -2, -0.35) -- (-2, -2, -0.25);
    \draw (2, -2, -0.35) -- (2, -2, -0.25);
    \draw (2, 2, -0.35) -- (2, 2, -0.25);
    
    \draw (-2, -2, -0.25) -- (2, -2, -0.25) -- (2, 2, -0.25);

    \draw [black,domain=0:0.4,variable=\x, smooth] plot ({-2, \x - 2, 0.125 * (-1 - cos(\x * 450))});
    \draw [black,domain=0:4,variable=\x, smooth] plot ({2, \x - 2, 0.125 * (-1 - cos(\x * 450))});

    \fill [white] (-2, -2, 0) -- (-2, 2, 0) -- (2, 2, 0) -- (2, -2, 0) -- cycle;
    \draw [dashed] (-2, -2, 0) -- (-2, 2, 0) -- (2, 2, 0) -- (2, -2, 0) -- cycle;

    \draw [opacity=0.3] (-2.5, 0, 0) -- (2.0, 0, 0);
    
    \begin{scope}[
      rotate around z = 20
      ]
      \draw [->] (-2.5, 0, 1) -- (0, 0, 0) node[pos=0,above] {$\vec{k}_{\mathrm{in}}$};
      \draw [opacity=0.3] (-2.5, 0, 0) -- (0, 0, 0);

      \draw [->] (0, 0, 0) -- (2.5, 0, 1) node[pos=1,above,align=center] {$\vec{k}_{\mathrm{out}}$ 
      };
      \draw [opacity=0.3] (0, 0, 0) -- (2.13, 0, 0);

      \draw [->] (0, 0, 0) -- (0.8, 0, 2) node[pos=1,above] {To SDD};

      \draw [opacity=0.5] [domain=180:158] plot ({1.8*cos(\x)},0,0 {1.8*sin(\x)},0,0);
      \draw (-1.65, 0, 0.9) node {$\theta$};

      \draw [opacity=0.5] [domain=0:22] plot ({1.8*cos(\x)},0,0 {1.8*sin(\x)},0,0);
      \draw (1.65, 0, 0.9) node {$\theta$};

    \end{scope}  

    \draw [opacity=0.5] [domain=180:200] plot ({2.3*cos(\x)}, {2.3*sin(\x)});
    \draw (-2.5, 0, 0) node[left] {$\phi$};

    \node[draw, rectangle, align=center, fill=black!20,
    rounded corners, align=center, left] at (-0.6, 0, 2) (fluo) {Fluorescence\\signal};
    
    \path[->] (fluo.south) edge[out=270] (-1.5,1,0.7);
    

    \foreach \t in {30,60,...,200}
    \foreach \f in {85,65,45,25}
    \draw [black, opacity=0.5, ->, thick]
    ({sin(\f - \h)*cos(\t - \h)}, {sin(\f - \h)*sin(\t - \h)}, {cos(\f - \h)})
    -- ({(1.1 + 0.2*cos(90 - \f))*sin(\f - \h)*cos(\t - \h)},
    {(1.1 + 0.2*cos(90 - \f))*sin(\f - \h)*sin(\t - \h)},
    {(1.1 + 0.2*cos(90 - \f))*cos(\f - \h)});

    \quadrant{200,210,...,390}{white}    
    
    \foreach \t in {230,260,...,390}
    \foreach \f in {85,65,45,25}
    \draw [black, opacity=0.5, ->, thick]
    ({sin(\f - \h)*cos(\t - \h)}, {sin(\f - \h)*sin(\t - \h)}, {cos(\f - \h)})
    -- ({(1.1 + 0.2*cos(90 - \f))*sin(\f - \h)*cos(\t - \h)},
    {(1.1 + 0.2*cos(90 - \f))*sin(\f - \h)*sin(\t - \h)},
    {(1.1 + 0.2*cos(90 - \f))*cos(\f - \h)});

  \end{scope}

\end{tikzpicture}
    \caption{Illustration of the GIXRF measurement process. X-ray light with
      wave vector $\vec{k}_{\rm in}$ is used to illuminate the sample. A
      fraction of the incoming light is first absorbed by the sample and
      afterwards emitted as a fluorescence spectrum. This can be detected by a
      silicon drift detector (SDD) which is oriented perpendicular to the
      incoming radiation, $\vec{k}_{\rm in}$. }
    \label{fig:gixrf_measurement}
\end{figure}

Grazing Incidence X-Ray Fluorescence (GIXRF) \cite{castor_gixrf,andrle2021shape}
is an indirect optical measurement method that can be used to quantify samples
both in terms of their geometry as well as in their material composition. To
investigate a sample, it is illuminated using X-ray light. The incident
radiation penetrates the sample to a certain depth that depends on the incoming
angle $\theta$ (c.f. \cref{fig:gixrf_measurement}). A large fraction of the
incoming radiation is reflected. The reflected radiation interferes with the
incident radiation and leads to the so-called X-ray standing wave (XSW) field. A
small fraction of the incident radiation is absorbed by the sample and promptly
given off again in the form of fluorescent light. Regions with constructive
interference contribute more strongly to the fluorescence signal. As penetration
depth and XSW depend strongly on the incidence angle $\theta$, so does also the
fluorescence spectrum, which is recorded by a calibrated silicon drift detector
(SDD) oriented perpendicular to the incident radiation beam $\vec{k}_{\rm
  in}$~\cite{H_nicke_2020}. To determine the geometrical (and experimental)
parameters of the sample using a set of experimentally obtained fluorescence
intensities, a parameterized forward model of the experimental measurement
process was created. In this model, a simulation of the electromagnetic fields
of the XSW is performed. The modified 2D Sherman equation is then used to
determine the numerical fluorescence intensity for each angle of incidence found
in the experimental dataset \cite{andrle2021shape}.

\begin{figure}[ht]
    \centering
    \input{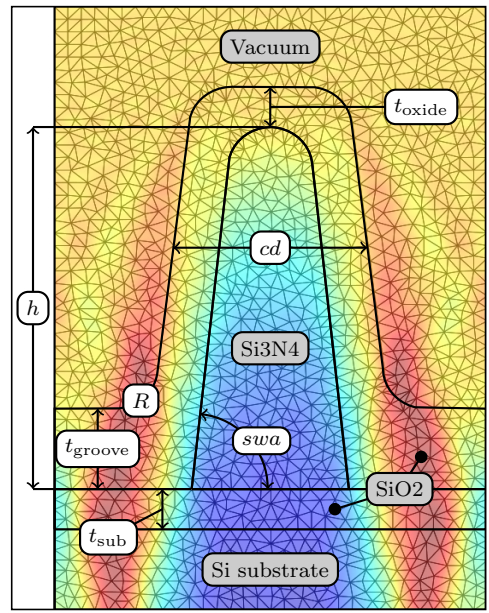}
    \caption{A unit cell of the GIXRF forward model with periodic boundary
      conditions in horizontal direction and perfectly absorbing boundary
      conditions in vertical direction. The geometric parameters are shown in
      white, the different materials are shown in gray. Additionally displayed
      are an exemplary finite element discretization and a simulated electric
      field intensity in pseudo-colors.}
    \label{fig:gixrf_overlay}
\end{figure}

\begin{figure}[ht]
    \centering
    \input{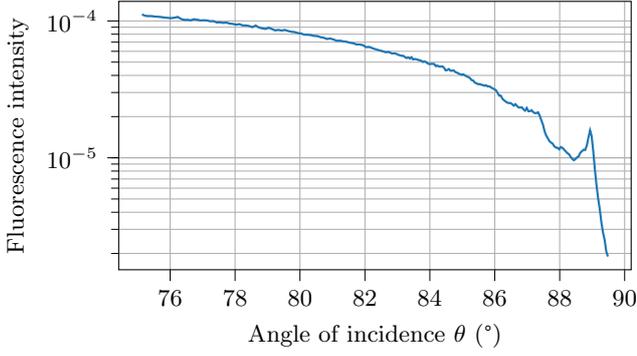}
    \caption{The experimentally recorded fluorescence intensity signal stemming
      from the Nitrogen in the core of the sample. The signal consists of
      \num{208} data points between angles of $\theta = \SI{75.13}{\degree}$ and
      $\theta = \SI{89.48}{\degree}$.}
    \label{fig:gixrf_signal}
\end{figure}

The experimental dataset obtained for X-ray light with an energy of
\SI{520}{\electronvolt} contains \num{208} discrete datapoints shown in
\cref{fig:gixrf_signal}. Note that the measurement uncertainties are
approximately two orders of magnitude smaller than the measurement signal. The
GIXRF reconstruction problem contains ten free parameters. Seven parameters
describe the shape of the sample (c.f. \cref{fig:gixrf_overlay}), one parameter
scales the calculated intensities to the experimental intensities, and two free
parameters have been introduced to account for uncertainties in the angles of
incidence $\theta$ and $\phi$ (c.f. \cref{fig:gixrf_measurement}). A listing of
the employed parameter optimization intervals is given in
\cref{tab:fitting_parameters_compact}. The densities of Si3N4 and SiO2,
$\varrho_{\mathrm{Si3N4}} = \SI{2.836}{\gram\per\centi\meter\cubed}$ and
$\varrho_{\mathrm{SiO2}} = \SI{2.093}{\gram\per\centi\meter\cubed}$,
were obtained by means of a separate X-ray
reflectometry experiment \cite{andrle2021shape}. The parameterized forward model
was created using the finite-element Maxwell solver JCMsuite.

\begin{table}
  \centering
  \begin{tabular}{ccc}
    \toprule
    Parameter & Range & Reconstruction results \\
    \midrule
    $h\, (\si{\nano\meter})$ & $\left[ 85, 100 \right]$ & \num{89.5 \pm 0.4} \\
    $cd\, (\si{\nano\meter})$ & $\left[ 35, 55 \right]$ & \num{46.3 \pm 0.2} \\
    $swa\, (\si{\degree})$ & $\left[ 75, 90 \right]$ & \num{83.7 \pm 0.1} \\
    $t_{\mathrm{oxide}}\, (\si{\nano\meter})$ & $\left[ 1, 6 \right]$ & \num{2.21 \pm 0.03} \\
    $t_{\mathrm{groove}}\, (\si{\nano\meter})$ & $\left[ 0.1, 10 \right]$ & \num{1.0 \pm 0.3} \\
    $t_{\rm Sub} \, (\si{\nano\meter})$ & $\left[ 0.1, 10 \right]$ & \num{6.9 \pm 0.9} \\
    $R\, (\si{\nano\meter})$ & $\left[ 3, 10 \right]$ & \num{7.0 \pm 0.8} \\
    $s_{\mathrm{N}}\, (1)$ & $\left[ 0.5, 1.5 \right]$ & \num{0.727 \pm 0.005} \\
    $\Delta_{\theta}\, (\si{\degree})$ & $\left[ -0.15, 0.15 \right]$ & \num{-0.101 \pm 0.003} \\
    $\Delta_{\phi}\, (\si{\degree})$ & $\left[ -0.075, 0.075 \right]$ & \num{0.006 \pm 0.009} \\
    \bottomrule
  \end{tabular}
  \caption{The fitting parameters, the corresponding ranges for the GIXRF model,
    as well as the results of the reconstruction due to the BTVO method, i.e.,
    $\vec{p}_{\rm LSQE} \pm \vecc{\epsilon}_{\rm LSQE}$. The reconstruction
    results were rounded to the first significant digit of the reconstructed
    uncertainties. Listed are the height $h$, the critical dimension $cd$, the
    side wall angle $swa$, the thickness of the oxide layer on top of the
    structure $t_{\mathrm{oxide}}$, the thickness of the oxide layer in the
    groove $t_{\mathrm{groove}}$, the thickness of the substrate $t_{\rm sub}$,
    the corner rounding radius $R$, a scaling parameter for the measured
    fluorescence $s_{\mathrm{N}}$, as well as offset values for the angle of
    incidence $\theta$ and the azimuth angle $\phi$, $\Delta_{\theta}$ and
    $\Delta_{\phi}$ respectively. }
  \label{tab:fitting_parameters_compact}
\end{table}

\subsubsection{Reconstruction results}
\label{subsec:gixrf_results}

\begin{figure*}[ht]
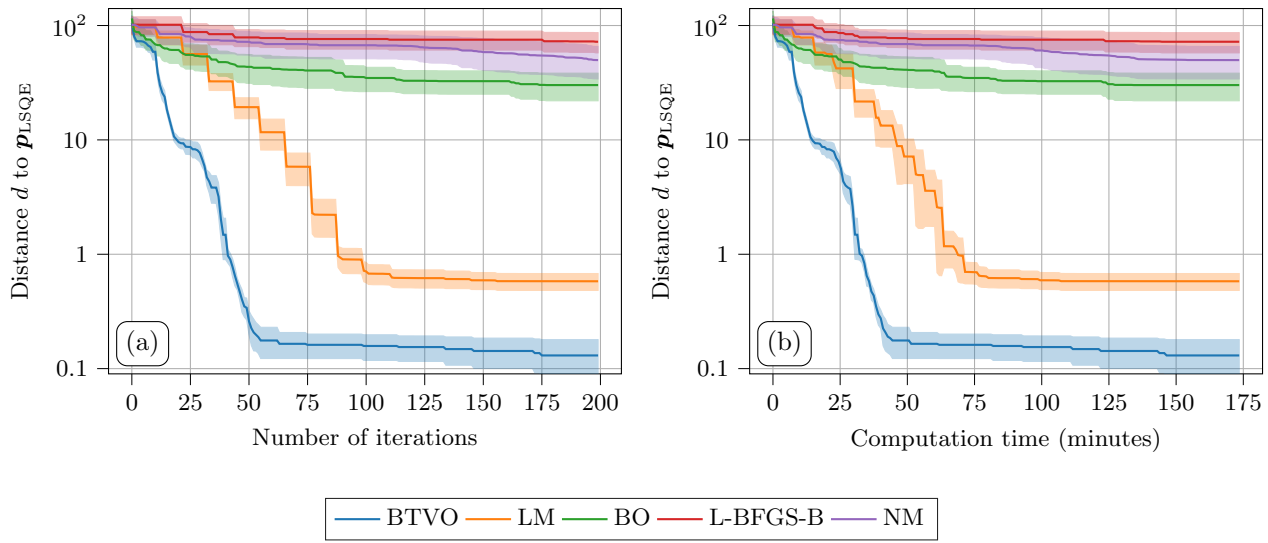

  \centering
  \begin{subfigure}[t]{0.49\textwidth}
  \vskip 0pt
    \centering
    \input{figure_ed2184d9f1f248e}
  \end{subfigure}
  \begin{subfigure}[t]{0.49\textwidth}
    \vskip 0pt
    \centering
    \input{figure_f4990a2ec5c2186}
  \end{subfigure}
  \definecolor{color0}{rgb}{0.12156862745098,0.466666666666667,0.705882352941177}
\definecolor{color1}{rgb}{1,0.498039215686275,0.0549019607843137}
\definecolor{color2}{rgb}{0.172549019607843,0.627450980392157,0.172549019607843}
\definecolor{color3}{rgb}{0.83921568627451,0.152941176470588,0.156862745098039}
\definecolor{color4}{rgb}{0.580392156862745,0.403921568627451,0.741176470588235}

\begin{tikzpicture}
    \begin{axis}[%
    font=\small,
    hide axis,
    xmin=10,
    xmax=20,
    ymin=0,
    ymax=0.1,
    legend style={draw=white!15!black,legend cell align=center,legend columns=-1}
    ]
    \addlegendimage{thick, color0}
    \addlegendentry{BTVO};
    \addlegendimage{thick, color1}
    \addlegendentry{LM};
    \addlegendimage{thick, color2}
    \addlegendentry{BO};
    \addlegendimage{thick, color3}
    \addlegendentry{L-BFGS-B};
    \addlegendimage{thick, color4}
    \addlegendentry{NM};
    \end{axis}
\end{tikzpicture}
  \caption{ The progress of the parameter reconstruction of the GIXRF dataset
    for different reconstruction methods. The plots show the distance
    $d(\vec{p})$ to the LSQE point $\vec{p}_{\rm LSQE}$, in units of the
    reconstructed Gaussian standard deviation $\vecc{\epsilon}_{\rm LSQE}$ (c.f.
    \cref{eq:distance}). Shown are the means (solid lines) and standard
    deviation (shaded bands) from six consecutive reconstruction runs. The x
    axis shows (a) the number of evaluations of the model function and (b) the
    time spent by each reconstruction method. None of the employed methods
    managed to reconstruct the parameters to an average distance of $d < 0.1$.
    Otherwise, the best reconstruction results are obtained by the BTVO scheme.
  }
  \label{fig:gixrf_reconstruction_results}
\end{figure*}

\cref{fig:gixrf_reconstruction_results} shows the performance of the employed
optimization methods at reconstructing the geometrical and experimental model
parameters for the GIXRF dataset. The benchmark's mean is depicted as a solid
line, while its standard deviation is shown as a shaded band around the mean.
\cref{fig:gixrf_reconstruction_results} (a) shows the progress of the
reconstruction in terms of calls to the model function, while
\cref{fig:gixrf_reconstruction_results} (b) shows it in terms of the actual wall
time. This differentiation highlights the slight computational overhead of the
Bayesian methods, where we are focusing on comparing the BTVO and LM in
particular. BTVO required approximately \num{40} iterations to cross the
(arbitrarily chosen) $d = 1$ threshold, while LM required approximately \num{88}
iterations, an increase of \SI{120}{\percent}. If we consider the wall time,
BTVO required approximately \SI{32}{\minute} to cross the $d = 1$ threshold and
LM required approximately \SI{68}{\minute}, which is a slightly smaller increase
of \SI{110}{\percent}. For model functions with very short computation times,
such as models described by analytic functions, BTVO may not necessarily be
advantageous. On average, a single forward simulation costs approximately
$\SI{30}{\second}$, while the generation of a new sample candidate with the BTVO
method costs on average between $\SI{2}{\second}$ and $\SI{3}{\second}$.
Dropping the approximation of using the same covariance kernel matrix for all
GPs would lead to an increase in the sample generation of over
$\SI{6}{\minute}$.

Of the employed reconstruction algorithms, only the proposed BTVO and LM managed
to reconstruct the parameters to an average of less than one standard deviation
of the LSQE. The BTVO reached an average distance of $d \approx \num{0.13}$,
while LM reached an average distance of $d \approx \num{0.58}$. The remaining
schemes did not manage to utilize the provided optimization budget to
reconstruct parameters within \num{10} standard deviations of the LSQE.

The reconstructed parameters and associated Gaussian uncertainties $\vec{p}_{\rm
  LSQE} \pm \vecc{\epsilon}_{\rm LSQE}$ due to the BTVO are found in
\cref{tab:fitting_parameters_compact}. The values were rounded to the first
significant digit of the reconstructed uncertainties. Explanations regarding to
the meaning of the parameters can be found in the table caption of
\cref{tab:fitting_parameters_compact}, as well as in
\cref{fig:gixrf_measurement,fig:gixrf_overlay}.

The X-ray light energy of \SI{520}{\electronvolt} is sufficient to obtain a
fluorescence signal from the Nitrogen within the core of the sample. Nevertheless,
also the different oxide layer thicknesses $t_{\rm oxide}$, $t_{\rm groove}$, 
and $t_{\rm sub}$ could be reconstructed with small uncertainty, which is in agreement
with the observation made by Soltwisch et al. \cite{C8NR00328A}.

\subsection{Analytical datasets: MGH17 and Gauss3}
\label{subsec:analytical}

\begin{table}[]
    \centering
    \begin{tabular}{ccc}
    \toprule
      Parameter & Range & Certified value \cite{NIST_MGH17} \\
      \midrule 
      $\beta_{1}$  & $[ \num{0}, \num{10}]$ & \num{0.374 \pm 0.002} \\
      $\beta_{2}$ & $[\num{0.1}, \num{4}]$ & \num{1.9 \pm 0.2} \\
      $\beta_{3}$ & $[\num{-4}, \num{-0.1}]$ & \num{-1.5 \pm 0.2} \\
      $\beta_{4}$ & $[\num{0.005}, \num{0.1}]$ & \num{0.0129 \pm 0.0004} \\
      $\beta_{5}$ & $[\num{0.005}, \num{0.1}]$ & \num{0.0221 \pm 0.0009} \\
      \bottomrule
    \end{tabular}    
    \caption{List of parameters with certified fit results as taken from
      \cite{NIST_MGH17}, as well as the employed parameter bounds for the MGH17
      dataset. The certified results are rounded to the first significant digit
      of the given uncertainty.}
    \label{tab:cert_vals_mgh17}
\end{table}

\begin{figure*}[ht]
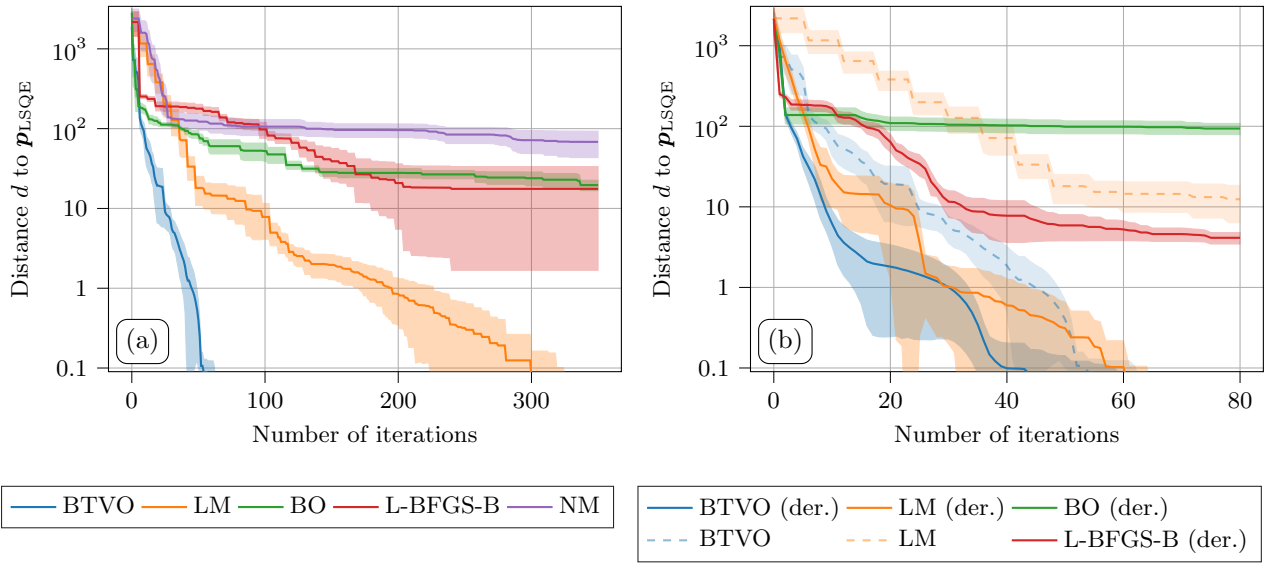

    \centering
    \begin{subfigure}[t]{0.49\textwidth}
        \vskip 0pt
        \centering
        \input{figure_5c9fa357eacdceb}
        \definecolor{color0}{rgb}{0.12156862745098,0.466666666666667,0.705882352941177}
\definecolor{color1}{rgb}{1,0.498039215686275,0.0549019607843137}
\definecolor{color2}{rgb}{0.172549019607843,0.627450980392157,0.172549019607843}
\definecolor{color3}{rgb}{0.83921568627451,0.152941176470588,0.156862745098039}
\definecolor{color4}{rgb}{0.580392156862745,0.403921568627451,0.741176470588235}

\begin{tikzpicture}
    \begin{axis}[%
    font=\small,
    hide axis,
    xmin=10,
    xmax=50,
    ymin=0,
    ymax=0.4,
    legend style={draw=white!15!black,legend cell align=center,legend columns=-1}
    ]
    \addlegendimage{thick, color0}
    \addlegendentry{BTVO};
    \addlegendimage{thick, color1}
    \addlegendentry{LM};
    \addlegendimage{thick, color2}
    \addlegendentry{BO};
    \addlegendimage{thick, color3}
    \addlegendentry{L-BFGS-B};
    \addlegendimage{thick, color4}
    \addlegendentry{NM};
    \end{axis}
\end{tikzpicture}
    \end{subfigure}
    \begin{subfigure}[t]{0.49\textwidth}
        \vskip 0pt
        \centering
        \input{figure_8e081125dadcda5}
        \definecolor{color0}{rgb}{0.12156862745098,0.466666666666667,0.705882352941177}
\definecolor{color1}{rgb}{1,0.498039215686275,0.0549019607843137}
\definecolor{color2}{rgb}{0.172549019607843,0.627450980392157,0.172549019607843}
\definecolor{color3}{rgb}{0.83921568627451,0.152941176470588,0.156862745098039}

\begin{tikzpicture}
    \begin{axis}[%
    font=\small,
    hide axis,
    xmin=10,
    xmax=50,
    ymin=0,
    ymax=0.4,
    legend style={draw=white!15!black,legend cell align=left,legend columns=3}
    ]
    \addlegendimage{thick, color0}
    \addlegendentry{BTVO (der.)};
    \addlegendimage{thick, color1}
    \addlegendentry{LM (der.)};    
    \addlegendimage{thick, color2}
    \addlegendentry{BO (der.)};
    \addlegendimage{thick, color0, dashed, opacity=0.5}
    \addlegendentry{BTVO};
    \addlegendimage{thick, color1, dashed, opacity=0.5}
    \addlegendentry{LM};
    \addlegendimage{thick, color3}
    \addlegendentry{L-BFGS-B (der.)};
  \end{axis}
\end{tikzpicture}
    \end{subfigure}
    \caption{The progress of the parameter reconstruction of the MGH17 dataset
      without (a) and with (b) the use of derivative information. Note that
      figure (b) shows a truncated view, limited to \num{80} iterations. The
      plots show the distance $d(\vec{p})$ to the LSQE point $\vec{p}_{\rm LSQE}$
      according to \cref{eq:distance}. Shown are the means (solid and dashed
      lines) and standard deviation (shaded bands) from six consecutive
      reconstruction runs. The $x$ axis shows the number of evaluations of the
      model function. Without the use of derivative information only BTVO and LM
      manage to reconstruct parameters that are within a distance $d <
      \num{0.1}$. Taking accurate derivative information into consideration
      improves both the BTVO and especially the LM results. However, BTVO still
      outperforms LM.}
    \label{fig:mgh17_res_iterations}
\end{figure*}

\begin{table}[]
  \centering
  \begin{tabular}{ccc}
    \toprule
    Parameter & Range & Certified value \cite{NIST_Gauss3} \\
    \midrule 
    $\beta_{1}$ & $[ \num{90}, \num{110}]$ & \num{98.9 \pm 0.5} \\
    $\beta_{2}$ & $[\num{0.005}, \num{0.05}]$ & \num{0.0109 \pm 0.0001} \\
    $\beta_{3}$ & $[\num{90}, \num{110}]$ & \num{100.7 \pm 0.8} \\
    $\beta_{4}$ & $[\num{100}, \num{120}]$ & \num{111.6 \pm 0.4} \\
    $\beta_{5}$ & $[\num{15}, \num{30}]$ & \num{23.3 \pm 0.4} \\
    $\beta_{6}$ & $[\num{70}, \num{80}]$ & \num{74 \pm 1} \\
    $\beta_{7}$ & $[\num{140}, \num{150}]$ & \num{147.8 \pm 0.4} \\
    $\beta_{8}$ & $[\num{17}, \num{22}]$ & \num{19.7 \pm 0.4} \\
    \bottomrule
  \end{tabular}    
  \caption{List of parameters with certified fit results as taken from
    \cite{NIST_Gauss3}, as well as the employed parameter bounds for the
    Gauss3 dataset. The certified results are rounded to the first significant
    digit of the given uncertainty.}
  \label{tab:cert_vals_gauss3}
\end{table}

\begin{figure*}[ht]
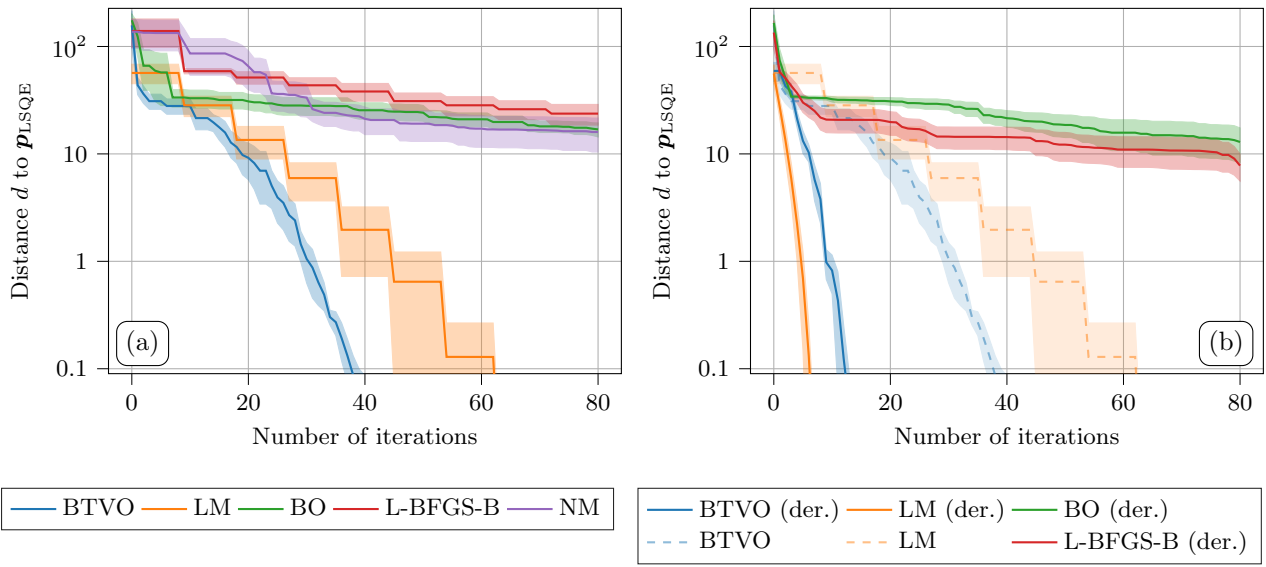

    \centering
    \begin{subfigure}[t]{0.49\textwidth}
        \vskip 0pt
        \centering
        \input{figure_f26e1f2db91ab0d}
        \definecolor{color0}{rgb}{0.12156862745098,0.466666666666667,0.705882352941177}
\definecolor{color1}{rgb}{1,0.498039215686275,0.0549019607843137}
\definecolor{color2}{rgb}{0.172549019607843,0.627450980392157,0.172549019607843}
\definecolor{color3}{rgb}{0.83921568627451,0.152941176470588,0.156862745098039}
\definecolor{color4}{rgb}{0.580392156862745,0.403921568627451,0.741176470588235}

\begin{tikzpicture}
    \begin{axis}[%
    font=\small,
    hide axis,
    xmin=10,
    xmax=50,
    ymin=0,
    ymax=0.4,
    legend style={draw=white!15!black,legend cell align=center,legend columns=-1}
    ]
    \addlegendimage{thick, color0}
    \addlegendentry{BTVO};
    \addlegendimage{thick, color1}
    \addlegendentry{LM};
    \addlegendimage{thick, color2}
    \addlegendentry{BO};
    \addlegendimage{thick, color3}
    \addlegendentry{L-BFGS-B};
    \addlegendimage{thick, color4}
    \addlegendentry{NM};
    \end{axis}
\end{tikzpicture}
    \end{subfigure}
    \begin{subfigure}[t]{0.49\textwidth}
        \vskip 0pt
        \centering
        \input{figure_e57c4a0615114dd}
        \definecolor{color0}{rgb}{0.12156862745098,0.466666666666667,0.705882352941177}
\definecolor{color1}{rgb}{1,0.498039215686275,0.0549019607843137}
\definecolor{color2}{rgb}{0.172549019607843,0.627450980392157,0.172549019607843}
\definecolor{color3}{rgb}{0.83921568627451,0.152941176470588,0.156862745098039}

\begin{tikzpicture}
    \begin{axis}[%
    font=\small,
    hide axis,
    xmin=10,
    xmax=50,
    ymin=0,
    ymax=0.4,
    legend style={draw=white!15!black,legend cell align=left,legend columns=3}
    ]
    \addlegendimage{thick, color0}
    \addlegendentry{BTVO (der.)};
    \addlegendimage{thick, color1}
    \addlegendentry{LM (der.)};    
    \addlegendimage{thick, color2}
    \addlegendentry{BO (der.)};
    \addlegendimage{thick, color0, dashed, opacity=0.5}
    \addlegendentry{BTVO};
    \addlegendimage{thick, color1, dashed, opacity=0.5}
    \addlegendentry{LM};
    \addlegendimage{thick, color3}
    \addlegendentry{L-BFGS-B (der.)};
  \end{axis}
\end{tikzpicture}
    \end{subfigure}
    \caption{ The progress of the parameter reconstruction of the Gauss3 dataset
      without (a) and with (b) the use of derivative information. The plots show
      the distance $d(\vec{p})$ to the LSQE point $\vec{p}_{\rm LSQE}$ according
      to \cref{eq:distance}. Shown are the means (solid and dashed lines) and
      standard deviation (shaded bands) from six consecutive reconstruction
      runs. The $x$ axis shows the number of evaluations of the model function.
      Both with and without derivative information, only BTVO and LM manage to
      reconstruct the parameters to within $d < \num{0.1}$ in the provided
      optimization budget. Without derivative information BTVO performs better
      than LM, and with derivative information LM performs better than BTVO.}
    \label{fig:gauss3_res_iterations}
\end{figure*}

Additionally, two analytical datasets were investigated. Here, the MGH17
\cite{NIST_MGH17} and Gauss3 \cite{NIST_Gauss3} datasets were chosen. Both were
obtained from the NIST Standard Reference Database \cite{NIST_StRD}. The target
values within the datasets were created using the respective model function,
where a normally distributed error was added to each data point. Since the model
functions are analytical in nature, derivatives are easily calculated. The model
function used to fit the 33 datapoints of the MGH17 dataset is
\begin{equation}
    \label{eq:mgh17_model}
    f(x,\vecc{\beta}) = \beta_{1} + \beta_2 \mathrm{e}^{-x \cdot \beta_4} + \beta_3 \mathrm{e}^{-x \cdot \beta_5}
\end{equation}
and contains 5 free parameters (c.f. \cref{tab:cert_vals_mgh17}). The Gauss3
dataset consists of \num{250} discrete datapoints and is fit using the 8
parameter model function (c.f. \cref{tab:cert_vals_gauss3})
\begin{equation}
    \label{eq:gauss3_model}
    f(x,\vecc{\beta}) = \beta_1 \mathrm{e}^{-\beta_2 \cdot x} + \beta_3 \mathrm{e}^{-(x-\beta_4)^2 / \beta_5^2 } + \beta_6 \mathrm{e}^{-(x-\beta_7)^2 / \beta_8^2} \,.
\end{equation}

\subsubsection{MGH17 reconstruction results}
\label{subsec:mgh17_results}

The results for the optimization benchmarks of the MGH17 dataset are shown in
\cref{fig:mgh17_res_iterations}. \cref{fig:mgh17_res_iterations} (a) shows the
results without the use of derivative information, while
\cref{fig:mgh17_res_iterations} (b) shows the results which were obtained with
accurate derivative information taken into consideration. For better visibility
of the convergence behavior of BTVO and LM, we have truncated the number of
iterations in \cref{fig:mgh17_res_iterations} (b) to \num{80}. The other methods
did not converge to $d < \num{0.1}$ within an optimization budget of \num{350}
iterations.

Without derivative information, only BTVO and LM were able to reconstruct the
certified results (c.f. \cref{tab:cert_vals_mgh17}) within its standard
deviations, as well as the uncertainty intervals. BTVO reached a value of $d <
0.1$ after approximately \num{54} iterations, while LM required \num{300}
iterations. None of the remaining methods were able to get to within 10 standard
deviations of $\vec{p}_{\rm LSQE}$ before the optimization budget was exhausted.

When derivative information was taken into consideration the results for BTVO,
LM, and L-BFGS-B improved. Both BTVO and LM were able to reach $d < 0.1$ (40
iterations for BTVO, 61 iterations for LM), and were ultimately also able to
reconstruct the certified result within the provided optimization budget. We
note that for this particular example BTVO \emph{without} derivative information
performs better than LM \emph{with} derivative information.

The performance of the conventional BO worsened when including derivative
information. We attribute this to the strong correlation of some of the
parameters for this specific reconstruction problem (c.f. MCMC sampling). This
is associated with large parameter regions with very similar values of $\chi^2$.
Due to the expected improvement infill criterion, BO has troubles to converge in
regions with very small gradients where it expects no relevant improvement. This
problem might be exacerbated when derivative information are available.

NM is not able to make use of derivative information and was therefore not
considered here.

\subsubsection{Gauss3 reconstruction results}
\label{subsec:gauss3_results}

The results for the optimization benchmarks of the Gauss3 dataset are shown in
\cref{fig:gauss3_res_iterations}, where \cref{fig:gauss3_res_iterations} (a)
depicts the results without the use of derivative information, while
\cref{fig:gauss3_res_iterations} (b) shows the results that were obtained with
accurate derivative information taken into consideration.

Without derivative information, BTVO again performed best and was able to
reproduce the values of the certified result (c.f. \cref{tab:cert_vals_gauss3})
and its uncertainty intervals. In each of the six runs it was able to
reconstruct parameters within less than \SI{10}{\percent} of the LSQE and reached
$d < \num{0.1}$ in approximately \num{38} iterations. LM was also able to
achieve $d < \num{0.1}$, and did so after \num{63} iterations. The remaining
three optimization schemes -- BO, L-BFGS-B, and NM -- were not able to get to
within 10 standard deviations of $\vec{p}_{\rm LSQE}$.

Taking derivative information into consideration improved the reconstruction
performance for all optimization schemes. L-BFGS-B was now also able to
reconstruct the model parameters to within \num{10} standard deviations. Very
drastic improvements were seen for both the proposed BTVO and LM schemes. BTVO
now only required approximately \num{13} iterations to reconstruct model
parameters within \SI{10}{\percent} of the LSQE, approximately a third of the
iterations required without derivative information. The clearest benefactor of
accurate derivative information however was LM, which was now able to
reconstruct the model parameters to within \SI{10}{\percent} in only \num{7}
iterations. The performance of BO increased slightly, although not to the extent
that we expected when providing accurate derivative information. We attribute
this again to the behavior described in \cref{subsec:mgh17_results}. However, as
the model parameters of the Gauss3 problem are correlated not as strongly as
those in the MGH17 problem, we can still observe a small improvement of the
reconstruction result. NM is again not considered, as it cannot make use of
derivatives.

\subsection{BTVO without effective degrees of freedom}
\label{sec:proper_dof_use}

As detailed in \cref{subsect:lsq_tvo}, keeping the DoFs fixed to the number of
data channels $K$ can lead to a very localized optimization behavior for the
BTVO scheme. The associated selection of samples that are too close to previous
sample positions leads to a termination of the optimization after a certain
number of iterations. We have observed that model functions with a smaller
number of data channels took longer to reach this point. The number of
iterations taken by each model function are averages from six different runs.
Here, the GIXRF reconstruction terminated after an average of \num{14}
iterations. The MGH17 reconstructions terminated after an average of \num{73}
iterations without, and \num{31} iterations with the use of derivatives. The
Gauss3 reconstructions terminated after an average of \num{14} iterations
without, and only \num{4} iterations with the use of derivatives. During this
small number of iterations, the BTVO scheme was only able to reconstruct
parameters with $d < 1$ for the MGH17 problem.

\subsubsection{Impact of the effective degrees of freedom on the
    approximated distribution}
\label{subsec:effdof_impact}

\begin{figure}[ht]
  \centering
  \centering
  \input{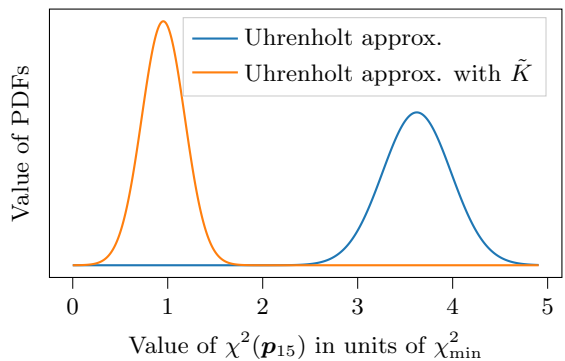}
  \caption{Illustrated is the impact of the effective DoF parameterization on
    the shape of the approximated probability distribution used by the employed
    infill criterion. For a discussion please refer to
    \cref{subsec:effdof_impact}.}
  \label{fig:effdof_pdf}
\end{figure}

By employing the effective DoF parameterization for the approximated probability
distribution, the parameter reconstruction of the GIXRF problem was successful
in each of the six instances in the benchmark \cref{subsec:gixrf_results}. When
fixing the effective DoFs to the full number of data channels $K$ we have
observed that the optimizations in the benchmark terminated on average after
\num{14} iterations, because the employed infill criterion was not able to
generate new valid samples.

We consider this particular iteration for an analysis of the impact of the
effective DoF parameterization. For a successful optimization we consider the
\emph{known} subsequent $\num{15}^{\rm th}$ iteration, i.e. its parameter
$\vec{p}_{15}$ and the associated observed model function values
$\vec{f}(\vec{p}_{15})$. As an example, we deliberately set the surrogate
uncertainty to be $\vec{\sigma} = 10 \cdot \vec{\eta}$ (\num{10} times the
measurement uncertainty) and calculate the approximated probability distribution
using an effective DoF parameterization and using a fixed $K$ DoF
parameterization, using \cref{eq:gammasq,eq:lambda} (c.f.
\cref{fig:effdof_pdf}).

We observe that both parameter distributions largely differ. While the mode of
the effective DoF probability distribution is close to the minimal observed
$\chi^2$ value, the probability density of the fixed $K$ DoF distribution is
almost zero for $\chi^2\approx\chi_{\rm min}^2$. Hence, the LCB infill criterion
in the fixed $K$ case does not choose the sample $\vec{p}_{15}$ but rather
parameters so close to the observed minimum of $\chi^2$ that the minimization is
stopped. This exemplifies, that the fixed $K$ parameterization hinders the
efficient exploration of the parameter space because it is too pessimistic about
finding small $\chi^2$ values.

\section{Application of Markov chain Monte Carlo sampling}
\label{subsec:mcmc_results}

\begin{figure*}[ht]
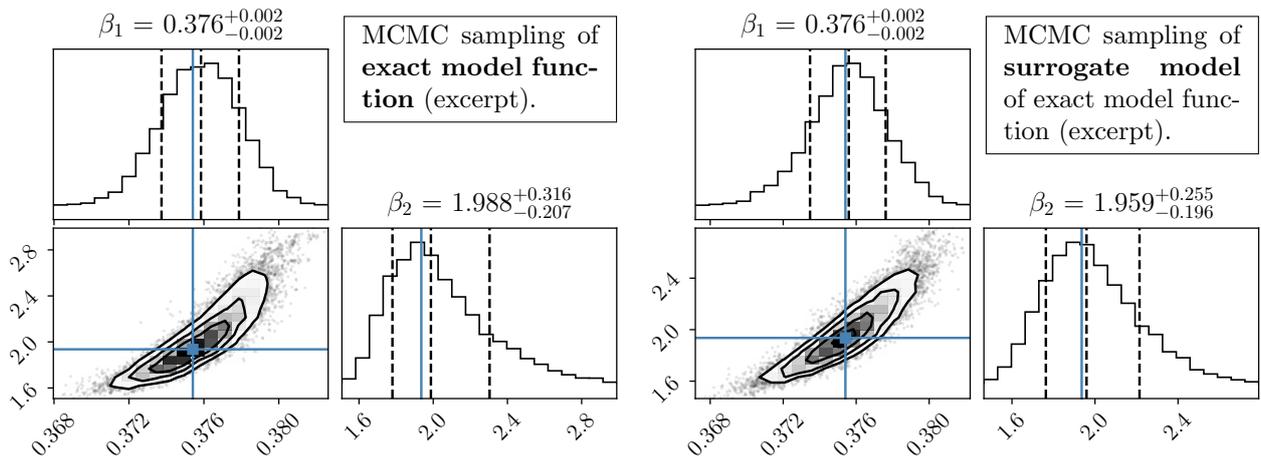

  \centering
  \begin{subfigure}[t]{0.49\textwidth}
    \vskip 0pt
    \centering
    \input{figure_6570af08dddca06}
  \end{subfigure}
  \begin{subfigure}[t]{0.49\textwidth}
    \vskip 0pt
    \centering
    \input{figure_1ad158ed354bfb6}
  \end{subfigure}
  \caption{An excerpt of the results of MCMC sampling of the likelihood function
    of the MGH17 dataset, using the exact likelihood function (left) directly
    and using a trained multi-output surrogate model (right) to calculate a
    stochastic prediction of the likelihood function. Only the parameters
    $\beta_1$ and $\beta_2$ are shown. A direct comparison of the two results
    reveals that the surrogate model approach is capable of reproducing the
    model parameter distributions as well as revealing correlations between the
    individual model parameters, while using significantly fewer evaluations of
    the model function. The blue lines shown are the certified NIST
    results~\cite{NIST_MGH17} (also c.f. \cref{tab:cert_vals_mgh17}). The
    remaining model parameters can be found in
    \cref{suppfig:mgh17_mcmc}\extfig.}
  \label{fig:mcmc_beta_1_2_comparison}
\end{figure*}

To demonstrate the capabilities of the surrogate model augmented MCMC sampling,
we applied it to the analytic MGH17 dataset. The goal was to sample the
predicted likelihood function $\hat{\mathcal{L}}(\vec{p})$, to extract the
parameter uncertainties in terms of the \SI{16}{\percent}, \SI{50}{\percent} (or
median), and \SI{84}{\percent} percentiles of the parameter distributions, and to
determine the correlations between the parameters of the model. Because the
MGH17 dataset is based on an analytic function, we also applied MCMC to the
exact likelihood function $\mathcal{L}(\vec{p})$ and compared the obtained
results. To sample the respective likelihood functions we employed
\texttt{emcee}~\cite{emcee}.

The surrogate model was trained in two stages. In the parameter reconstruction
stage the LSQE was found within \num{63} iterations. Derivative information was
not used during the reconstruction. In the refinement stage, the uncertainty of
the surrogate model in the region of interest around the LSQE is reduced. At each
refinement step, $S = 10 \cdot (N + 1)$ random samples were drawn and the one
with the largest mean uncertainty was used to evaluate the model function and
retrain the surrogate model. The refinement was stopped after the maximum mean
uncertainty was below $\sigma_{\rm min} = \num{1e-4}$ times that of the global
uncertainty in the surrogate model for five consecutive iterations, which was
reached after an additional \num{41} iterations. The actual model function was
thus evaluated \num{104} times.

\cref{fig:mcmc_beta_1_2_comparison} shows an excerpt of the comparison of MCMC
sampling of the exact likelihood function $\mathcal{L}(a, b, \vec{p})$ and the
predicted likelihood function $\hat{\mathcal{L}}(a, b, \vec{p})$ around the LSQE
$\vec{p}_{\rm LSQE}$. In both instances, 32 MCMC walkers were used to draw
\num{50000} samples from the respective likelihood functions. The complete
comparison for all parameters can be found in \cref{suppfig:mgh17_mcmc}\extfig.

The histograms at the top show the \SI{16}{\percent}, \SI{50}{\percent} (or
median), and \SI{84}{\percent} percentiles of the parameter distributions. Both
the exact, as well as the surrogate aided MCMC samplings produced very similar
histograms, both in terms of percentile values as well as in terms of the
overall shape. The recovered values matched the certified results provided by
the NIST within the reconstructed standard deviations, see also
\cref{tab:cert_vals_mgh17}.

The 2d scatter plots between the histograms highlight correlations between the
model parameters. The displayed levels are -- from inside to outside -- the
$0.5\sigma$, $1\sigma$, $1.5\sigma$, $2\sigma$ regions. Again, both methods
showed good agreement.

Finally, the method was used to assess the magnitude of the measurement
uncertainties. To this end the basic error model
\begin{equation*}
  \label{eq:mcmc_error_model}
  \tilde{\eta}_i = 2^{\eta_c} \cdot \eta_i
\end{equation*}
was fit to the data. The actual measurement uncertainties can be scaled up and
down by fitting the value of the exponent $\eta_c$. Both MCMC sampling of the
exact model and the surrogate model yield a value of $\eta_c$ close to zero with
an uncertainty of about $\pm 0.2$, c.f. \cref{suppfig:mgh17_mcmc}\extfig.

This demonstrates that qualitatively and quantitatively accurate results can be
obtained with the surrogate model aided MCMC, which comes at a fraction of the
cost of directly sampling the true likelihood function if the model function is
expensive to evaluate.

\subsection{Surrogate augmented MCMC convergence behavior}
\label{subsec:surr_mcmc_conv}

\begin{figure}[h]
  \centering
\begin{tikzpicture}

\begin{axis}[
font=\small,
grid=both,
height=0.225 \textheight,
legend cell align={left},
legend style={fill opacity=0.8, draw opacity=1, text opacity=1, at={(0.97,0.97)}, anchor=north east, draw=white!80!black},
legend style={fill opacity=0.8, draw opacity=1, text opacity=1, draw=white!80!black},
log basis y={10},
tick align=outside,
tick pos=left,
width=0.95 \columnwidth,
x grid style={white!69.0196078431373!black},
xlabel={Added sampling points after optimization},
xmajorgrids,
xmin=-7.5, xmax=157.5,
xtick style={color=black},
xtick={0, 10, 20, 30, 40, 50, 75, 100, 150},
y grid style={white!69.0196078431373!black},
ylabel={MCMC percentile avg. rel. err.},
ymajorgrids,
ymin=0.0075, ymax=0.055,
ymode=log,
ytick style={color=black},
ytick={8e-3, 1e-2, 2e-2, 3e-2, 4e-2, 5e-2},
yticklabels={\(\displaystyle {8 \cdot 10^{-3}}\),\(\displaystyle {10^{-2}}\),\(\displaystyle {2 \cdot 10^{-2}}\),\(\displaystyle {3 \cdot 10^{-2}}\),\(\displaystyle {4 \cdot 10^{-2}}\),\(\displaystyle {5 \cdot 10^{-2}}\),}
]
\addplot [semithick, black, mark=*, mark size=3, mark options={solid}, only marks, forget plot]
table {%
0 0.0451510530256331
5 0.0351405885897276
10 0.0343717328798397
20 0.0309071938175221
30 0.0227971561477304
40 0.0221474329977523
50 0.0180388519919311
75 0.0218298446940652
100 0.0180561080952446
150 0.0137750152707538
};
\addplot [thick, black]
table {%
0 0.0342098908177996
1.5 0.0338619774280466
3 0.0335176023052647
4.5 0.0331767294654031
6 0.0328393232903673
7.5 0.0325053485242975
9 0.0321747702698849
10.5 0.0318475539847252
12 0.0315236654777088
13.5 0.0312030709054489
15 0.0308857367687445
16.5 0.0305716299090803
18 0.030260717505162
19.5 0.0299529670694868
21 0.0296483464449485
22.5 0.0293468238014776
24 0.0290483676327153
25.5 0.0287529467527213
27 0.0284605302927155
28.5 0.0281710876978518
30 0.027884588724026
31.5 0.0276010034347156
33 0.027320302197851
34.5 0.0270424556827202
36 0.0267674348569032
37.5 0.0264952109832387
39 0.0262257556168216
40.5 0.0259590406020301
42 0.0256950380695845
43.5 0.0254337204336343
45 0.0251750603888763
46.5 0.0249190309077014
48 0.0246656052373702
49.5 0.0244147568972176
51 0.0241664596758863
52.5 0.0239206876285873
54 0.0236774150743891
55.5 0.0234366165935348
57 0.023198267024785
58.5 0.0229623414627896
60 0.022728815255485
61.5 0.0224976640015181
63 0.0222688635476967
64.5 0.0220423899864658
66 0.0218182196534094
67.5 0.0215963291247777
69 0.0213766952150397
70.5 0.0211592949744603
72 0.0209441056867027
73.5 0.0207311048664541
75 0.0205202702570767
76.5 0.020311579828282
78 0.0201050117738285
79.5 0.0199005445092438
81 0.0196981566695685
82.5 0.0194978271071242
84 0.0192995348893038
85.5 0.0191032592963837
87 0.0189089798193595
88.5 0.0187166761578026
90 0.0185263282177389
91.5 0.0183379161095495
93 0.0181514201458921
94.5 0.017966820839644
96 0.0177840989018659
97.5 0.0176032352397863
99 0.0174242109548065
100.5 0.0172470073405259
102 0.0170716058807873
103.5 0.0168979882477424
105 0.0167261362999361
106.5 0.0165560320804115
108 0.0163876578148334
109.5 0.0162209959096307
111 0.0160560289501584
112.5 0.015892739698878
114 0.015731111093556
115.5 0.0155711262454812
117 0.0154127684377001
118.5 0.0152560211232699
120 0.0151008679235298
121.5 0.0149472926263893
123 0.0147952791846343
124.5 0.0146448117142503
126 0.0144958744927626
127.5 0.0143484519575937
129 0.0142025287044368
130.5 0.0140580894856465
132 0.0139151192086452
133.5 0.0137736029343466
135 0.0136335258755941
136.5 0.0134948733956161
138 0.0133576310064965
139.5 0.0132217843676606
141 0.0130873192843771
142.5 0.0129542217062743
144 0.0128224777258727
145.5 0.012692073577131
147 0.0125629956340083
148.5 0.0124352304090399
150 0.0123087645519282
};
\addlegendentry{Least-square trend line}
\path [draw=black, very thick, dash pattern=on 5.55pt off 2.4pt]
(axis cs:0,0.00794937003851718)
--(axis cs:150,0.00794937003851718);
\addlegendimage{draw=black, very thick, dash pattern=on 5.55pt off 2.4pt}
\addlegendentry{Theoretical threshold}

\end{axis}

\end{tikzpicture}
  \caption{The average relative deviation from the exact MCMC percentiles for
    different numbers of points added during a refinement of the surrogate of
    the MGH17 model function. For a discussion c.f.
    \cref{subsec:surr_mcmc_conv}.}
  \label{fig:mcmc_convergence}
\end{figure}
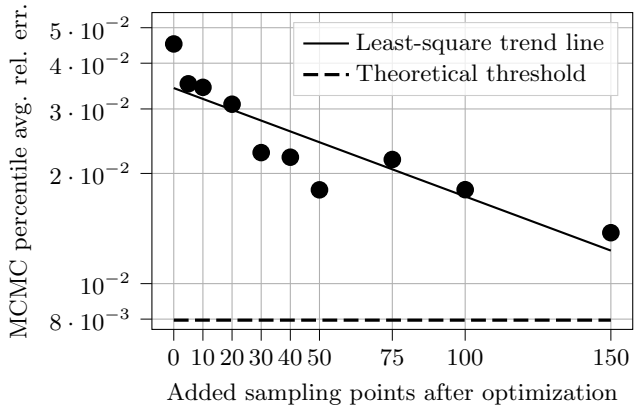

The accuracy of the surrogate augmented MCMC sampling depends strongly on the
accuracy of the underlying GP model and therefore on the number of additional
training samples for the GP evaluated after the optimization. In order to
illustrate the convergence of the MCMC results with respect to the number of
additional samples, we considered the following setup. We start from the MGH17
optimization results from \cref{subsec:mcmc_results} and consider refinements of
the GP surrogate with $N\in{0,5,10,20,30,40,50,75,100,150}$ additional training
samples. Using the refined surrogate we then drew \num{500000} samples using 32
MCMC walkers. From these samples we determined the \SI{16}{\percent},
\SI{50}{\percent}, and \SI{84}{\percent} percentiles for each parameter
$\beta_{i}$. To create reference values we employed \texttt{emcee} to draw
\num{500000} samples from the exact likelihood function for six consecutive
runs. For each run we then determined the same percentiles as described above,
the reference percentiles were then obtained by calculating the mean values
across the six runs. The number of samples drawn was increased to improve the
stability of the determined percentiles. The relative deviation from the exact
likelihood result was determined for each percentile and parameter $\beta_{i}$
and then averaged for each $N$ across these parameters and percentiles. The
results are shown in \cref{fig:mcmc_convergence}. To better highlight the trend,
a least-squares fit trend line is shown as well. A theoretical threshold of
$\overline{\epsilon}_{\rm rel, MCMC perc.} \approx \num{8e-3}$ was determined by
determining the average relative standard deviation from the six reference value
runs.

We observe that for an increasing number of refinement samples $N$ the relative
deviation from the exact results decreases from initially \SI{5}{\percent} to
about \SI{1}{\percent}. This shows that accurate uncertainty estimates can be
obtained with a relatively small number of forward model evaluations.

\subsection{MCMC sampling of the approximate likelihood function of the
    experimental GIXRF dataset}
\label{subsec:gixrf_mcmc}

\begin{figure}[h]
  \centering
  \includegraphics[width=\columnwidth]{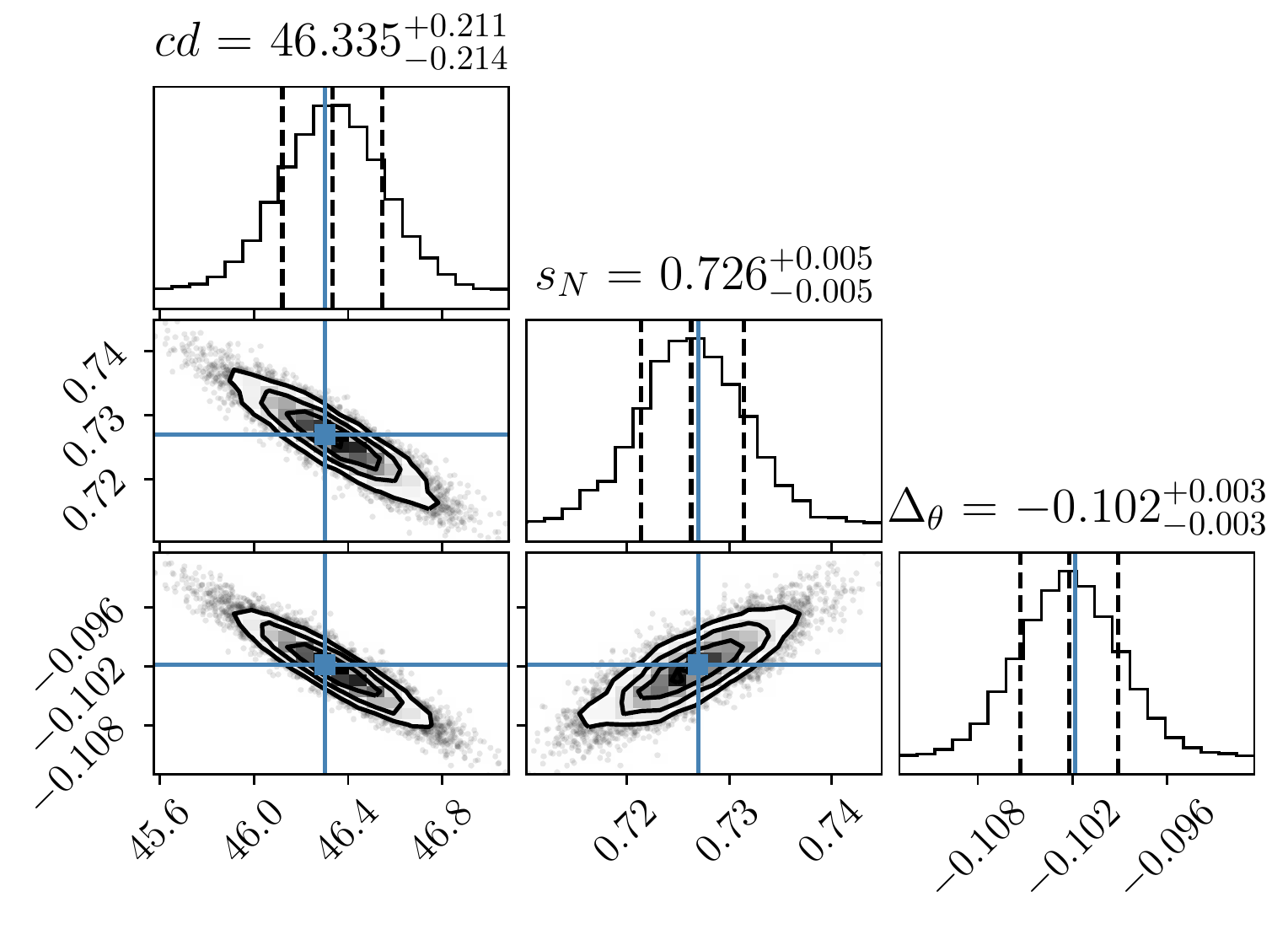}
  \caption{The approximate likelihood function of the GIXRF dataset was sampled
    using the surrogate augmented MCMC method. The displayed parameters are
    limited to the critical dimension $cd$, the scaling parameter $s_{N}$, and
    the offset of the incidence angle $\theta$. For these parameters we observe
    correlations at the reconstructed parameter value $\vec{p}_{\rm LSQE}$. The
    blue lines are the results of the parameter reconstruction. The results
    shown in the histograms confirm the results from the BTVO reconstruction,
    c.f. \cref{tab:fitting_parameters_compact}.}
  \label{fig:gixrf_mcmc}
\end{figure}

Using the surrogate augmented MCMC method we have sampled the approximate
likelihood function of the GIXRF dataset. An excerpt of the results is shown in
\cref{fig:gixrf_mcmc}, where we have limited the displayed parameters to the
critical dimension $cd$, the scaling parameter $s_{N}$, and the offset of the
incidence angle $\theta$. For these parameters we observe correlations at the
reconstructed parameter value $\vec{p}_{\rm LSQE}$. This is due to the fact that
the right-most maximum in the measurement data, c.f. \cref{fig:gixrf_signal} at
approximately \SI{89}{\degree}, is strongly correlated to the critical dimension
$cd$ \cite{C8NR00328A}. A change in the offset angle $\Delta_{\theta}$ moves the
peak, which leads to a correlation between $cd$ and $\Delta_{\theta}$. The
correlation between $s_N$ and $cd$ can be explained if we consider, that a
larger critical dimension means that the grating contains more material, which
increases the fluorescence signal. $s_N$ is the parameter that scales this
signal to match the experimentally observed one. In order to still match the
experimental measurement when increasing the critical dimension, the
fluorescence signal has to be scaled down. The results shown in the histograms
confirmed the results obtained from the reconstruction using the BTVO method,
c.f. \cref{tab:fitting_parameters_compact}.
  
The remaining model parameters showed only minor correlations.

\section{Conclusion}
\label{sec:discussion}

We have expanded a recently introduced Bayesian target-vector optimization
(BTVO) scheme to be compatible with a much larger number of data channels $K$.
We have done this by introducing a new parameterization, which adjusts the
actual DoFs to be an effective number of DoFs, as
well as by sharing the covariance function for the individual GPs among each
other, such that only one matrix decomposition is necessary at each iteration of
the optimization instead of one for each data channel.

We have applied the proposed BTVO to various parameter reconstruction tasks and
have shown that the scheme regularly outperforms the established
Levenberg-Marquardt algorithm at reconstructing model parameters during a
least-squares fit, the single exception being when derivative information has
been directly exploited in one particular example.

We have further shown that the method can be extended to enable Markov chain
Monte Carlo sampling of expensive model functions with largely reduced
computational costs. This has been achieved by exploiting the multi-output
surrogate model which was trained during the parameter reconstruction. By using
an analytic model function we have demonstrated that the surrogate model
approach is capable of generating results that are in good agreement with the
exact results, while using only a fraction of the evaluations of the model
function. In doing this, one is able to obtain model parameter uncertainties in
terms of percentiles of the model parameter distributions, to determine
non-linear correlations between parameters, and in principle also to fit a
non-trivial error model to the data.

\section{Supporting information}

Supporting information can be found in the supplementary material.

\section{Conflict of Interest}

The authors declare no conflict of interest.

\section{Acknowledgements}
\label{sec:acknowledgements}

We acknowledge discussions with Victor Soltwisch, Martin Hammerschmidt, and Lin
Zschiedrich, and we acknowledge Philipp Hönicke for assistance in recording the
experimental data set. This project is funded by the German Federal Ministry of
Education and Research (BMBF, project number 05M20ZAA, siMLopt; project number
01IS20080A, SiM4diM; Forschungscampus MODAL, project number 05M20ZBM), by the
German Federal Ministry for Economic Affairs and Energy (BMWi, project number
50WM2067, Optimal-QT), and by the European Union's Horizon 2020 research and
innovation programme (EU H2020, grant number 101007319, AI-TWILIGHT). This
project has received funding from the EMPIR programme co-financed by the
Participating States and from the European Union’s Horizon 2020 research and
innovation programme (project 20IND04 "ATMOC"; project 20FUN02 "POLIGHT").

\bibliographystyle{MSP}
\bibliography{bibl}

\newpage

\setcounter{figure}{1}
\setcounter{equation}{1}
\setcounter{section}{1}
\setcounter{page}{1}

\renewcommand{\thefigure}{S\arabic{figure}}
\renewcommand{\theequation}{S\arabic{equation}}
\renewcommand{\thesection}{S\arabic{section}}
\renewcommand{\thepage}{S\arabic{page}}

\onecolumn
\appendix
\part*{Supplementary Material}
\setcounter{equation}{0}
\setcounter{figure}{0}
\renewcommand{\theequation}{S\arabic{equation}}
\renewcommand{\thefigure}{S\arabic{figure}}

\section{Parameter reconstruction using Bayesian inference}
\label{suppsec:bayesian_param_recon}

In this section we go into more detail on the probabilistic background of
parameter reconstructions. Given some experimental measurement $\vec{t} = (t_1,
\dots, t_K)\trans$ and a parameterized vectorial model $\vec{f}(\vec{p})$ of
that measurement with $\vec{p} \in \mathcal{X} \subset \mathbb{R}^{N}$ and
$\vec{f}: \mathcal{X} \to \mathbb{R}^{K}$, a parameter reconstruction consist of
finding model parameters $\vec{p}$ such that $\vec{f}(\vec{p})$ best matches
$\vec{t}$. For the true parameter vector $\vec{p}_{t}$ the $i$-th model output
is equal to the measurement result plus a noise contribution, i.e.
\begin{equation*}
  \label{suppeq:measurement}
  t_i = f_i(\vec{p}_t) + \epsilon_i \,.
\end{equation*}
We further assume no correlations between the measurements and no systematic
errors. The noise is modeled to be normally distributed with zero 
mean and variance $\eta_i^2$, i.e. $\epsilon_i\sim\mathcal{N}(0,\eta_i^2)$.

To obtain an estimate for the true parameter in the Bayesian sense one
defines a probability density function (PDF) for the model parameters and
maximizes this density with respect to all parameters of the density. 
A suitable PDF for the reconstruction problem is the likelihood function 
for $K$ independent normally distributed random variables~\cite{degroot2012probability},
\begin{equation*}
  \label{suppeq:likelihood_function_1}
  \mathcal{L}(\vec{p}) = \prod_{i=1}^K \frac{1}{\sqrt{2\pi
      \eta_i^2}}\exp\left(-\frac{1}{2} \frac{\left(f_i(\vec{p}) -
        t_i\right)^2}{\eta_i^2}\right) \,.
\end{equation*}
The maximum likelihood point estimator (MLE) is defined as
\begin{equation*}
  \label{suppeq:mle}
  \vec{p}_{\rm MLE} = \underset{\vec{p}}{\rm arg\,max}\, \mathcal{L}(\vec{p}) \,.
\end{equation*}
In cases where prior knowledge about the distribution of the model parameter
exists, such as for example geometrical parameter constraints~\cite{Hamm:17},
one may also choose to instead maximize the full posterior probability density
\begin{equation*}
  \label{suppeq:posterior_pd}
  \mathcal{P}(\vec{p}) \propto \pi(\vec{p})\mathcal{L}(\vec{p}) \,,
\end{equation*}
which is -- up to a constant of proportionality -- obtained by multiplying the
prior probability density $\pi(\vec{p})$ with the likelihood function. The
maximum a-posteriori point estimator (MAP)
\begin{equation*}
  \label{suppeq:map}
  \vec{p}_{\rm MAP} = \underset{\vec{p}}{\rm arg\,max}\, \mathcal{P}(\vec{p})
\end{equation*}
or the MLE are generally determined by means of a local optimization using
methods such as the Nelder-Mead simplex algorithm or the gradient based L-BFGS-B
method \cite{Hamm:17}. If the probability distribution exhibits multiple local
minima, global heuristic optimization such as particle swarm optimization are
employed \cite{Solt:2017}. In the context of scatterometry, it is often
expensive to evaluate the model function. In this case, Bayesian optimization
methods often significantly reduce the computational
effort~\cite{Schn:2019Benchmark,Schn:2019GPR}.

Often, the magnitude of the variance $\eta_i^2$ is \emph{a priori} unknown.
However, it is often possible to define a realistic model for the variance and
determine the most likely model parameters. A possible approach is to model the
measurement variances for example as~\cite{Henn:12}
\begin{equation*}
  \label{suppeq:common_error_model}
  \eta_i^2(a,b,\vec{p}) = \left(a \cdot f_i(\vec{p})\right)^2 + b^2 \,,
\end{equation*}
where one assumes that the error is composed of a background term $b$ and a
linear dependent noise, which could, e.g., stem from intensity-proportional
power fluctuations of laser light. This can be incorporated into the parameter
reconstruction simply by extending the parameter space of the respective PDFs,
i.e. by considering the likelihood function
\begin{equation*}
  \label{suppeq:likelihood_function_error_model}
  \mathcal{L}(a, b, \vec{p}) = \prod_{i=1}^K \frac{1}{\sqrt{2\pi
      \eta_i^2(a, b, \vec{p})}}\exp\left(-\frac{\left(f_i(\vec{p}) -
        t_i\right)^2}{2\eta_i^2(a, b, \vec{p})}\right) \,.
\end{equation*}

An important information in the context of parameter reconstruction are not only
the point estimates but also the confidence intervals of the parameter values.
These are often given as the $\SI{16}{\percent}$, the $\SI{50}{\percent}$
(median), and the $\SI{84}{\percent}$ percentiles of the probability density
distributions $\mathcal{P}$ or $\mathcal{L}$. The percentiles can generally not
be calculated in closed form, but are determined by drawing samples from the
probability distribution using Markov chain Monte Carlo (MCMC) sampling
techniques~\cite{Solt:2017}. For stable percentile estimates often more than
10,000 samples are required.

\section{Weighted least-squares}
\label{suppsec:levenberg-marquardt}

The problem of finding the MAP or MLE, as defined in
\cref{suppsec:bayesian_param_recon}, can in principle be transformed into that
of solving a weighted least-square problem. Two assumptions go into this. First,
when attempting to find the MAP the prior is sufficiently flat, such that it may
be neglected. The logarithm of the posterior probability density
$\mathcal{P}(\vec{p})$ is then up to a constant addition equal to the logarithm
of the likelihood $\mathcal{L}(\vec{p})$, i.e. $\log \mathcal{P}(\vec{p}) =
\mathrm{const.} + \log\mathcal{L}(\vec{p})$. And second, a good approximation of
the individual error variances exists. We assume that the relative error
variances $\eta_i^2 / \eta_j^2$ are known, but all of them may be over- or
underestimated. We can then write
\begin{equation}
  \label{suppeq:posterior_chi_sq}
  -2 \log\mathcal{P}(\vec{p}) + \mathrm{const.} = \sum_{i=1}^K
  \frac{\left(f_i(\vec{p}) - t_i\right)^2}{\eta_i^2} = \left(\vec{f}(\vec{p})
  - \vec{t}\right)^{\rm T} \mat{W} \left(\vec{f}(\vec{p}) - \vec{t}\right) =
  \chi^2(\vec{p}) \,,
\end{equation}
where $\mat{W} = \diag(1/\eta_1^2,\dots,1/\eta_K^2)$ is a diagonal matrix
containing the error variances. Solving the least-square problem, i.e. finding
the $\vec{p}$ that minimizes \cref{suppeq:posterior_chi_sq}, is therefore
equivalent to finding the MAP or MLE under the assumptions stated above. This
can be efficiently done using, e.g., the Gauss-Newton method
\cite{deuflhard2005newton,seber2003nonlinear} or the Levenberg-Marquardt
algorithm \cite{leve:1944,marquardt1963algorithm,seber2003nonlinear}.

The approaches work by determining at each iteration $m$ at position $\vec{p}_m$
a linear approximation of the model function
\begin{equation}
  \label{suppeq:linear_approx_f}
  \vec{f}(\vec{p}_m+\vec{\delta}) \approx \vec{f}(\vec{p}_m) + \mat{J}\vec{\delta} \,,
\end{equation}
where $\mat{J}$ is the Jacobian matrix with entries $\mat{J}_{ij}= \partial
f_i(\vec{p}) / \partial p_j |_{\vec{p} = \vec{p}_m}$. The derivative of $\chi^2$
with respect to the step size $\vec{\delta}$ vanishes in the linear
approximation for $\vec{\delta}_{\rm GN}$ fulfilling
\begin{equation*}
  \left (\mat{J}\trans \mat{W} \mat J \right
  )\vec{\delta}_{\rm GN} = \mat{
    J}\trans \mat{W} \left [\vec{t} -
    \vec{f}(\vec{p}_k)\right ] \,.
\end{equation*}
This Gauss-Newton step $\vec{p}_{m+1} = \vec{p}_{m} + \vec{\delta}_{\rm GN}$
can be misleading, if it is larger than the validity range of the linear
approximation of $\vec{f}(\vec{p})$. A more conservative approach is to make a
small step in the direction of the gradient $\nabla \chi^2(\vec{p})$. The
Levenberg-Marquardt algorithm~\cite{leve:1944} with improvements by R.
Fletcher~\cite{flet:1971} aims to combine both strategies by introducing a
damping factor $\lambda$ that is adjusted according to the success of previous
optimization steps and solving the equation
\begin{equation*}
  \label{suppeq:LM}
  \left(
    \mat{J}\trans \mat{W} \mat J
    + \lambda \cdot \diag [ \mat{J}\trans \mat{W} \mat{J} ]
  \right) \vec{\delta}_{\rm LM}
  = \mat{J}\trans \mat{W} \left[\vec{t} - \vec{f}(\vec{p}_k) \right]
  \,.
\end{equation*}

The Levenberg-Marquard algorithm is typically able to find the minimum of
$\chi^2$ with fewer iterations than other local optimization methods like the
Nelder-Mead downhill-simplex approach or L-BFGS-B. This is because the
Levenberg-Marquardt algorithm uses a linear model of \emph{each} channel
$f_i(\vec{p})$ resulting in an accurate local \emph{second-order} model of
$\chi^2$. On the other hand, L-BFGS-B uses an accurate local first-order model
of $\chi^2$ and only builds up an averaged approximation of the second-order
Hessian matrix $(\mat{H})_{ij} = \frac{1}{2}\frac{\partial^2 \chi^2}{\partial
  p_i \partial p_j}$ during the minimization. It can be therefore beneficial to
solve the weighted least square problem instead of maximizing the full posterior
probability distribution $\mathcal{P}(\vec{p})$.

Under the assumption that the linear approximation of \cref{suppeq:linear_approx_f}
is valid in a sufficiently large region around the minimum of $\chi^2$, it is
possible to determine the confidence intervals of each parameter analytically by
the diagonal elements of the parameter covariance matrix $\mat{Cov}(\vec{p}_{\rm
  MLE})
= \mat{H}^{-1} \approx (\mat{J}\trans \mat{W}
\mat{J})^{-1}$~\cite{strutz2010data,press2007numerical}. A possible over or
underestimation of the measurement errors $\eta_1,\dots,\eta_K$ in the weight
matrix $\mat{W}$ can be accounted for by a scaling with the regression standard
error~\cite{strutz2010data,kutner2005applied}
\begin{equation*}
  \label{suppeq:rse}
  {\rm RSE} = \sqrt{\frac{\chi^2(\vec{p}_{\rm
        MLE})}{K-N}} \,.
\end{equation*}
Recall that the regression standard error can be used to indicate how well the
error variances were estimated. Assuming a correct model ${\rm RSE} < \num{1}$
indicates that the measurement uncertainties were overestimated and ${\rm RSE} >
\num{1}$ that the measurement uncertainties were underestimated. The measurement
uncertainties for the parameters $p_i$ for $i=1,\dots,N$ evaluate
to
\begin{equation*}
  \label{suppeq:lm_uncertainties}
  \epsilon_{p_i} = {\rm RSE}\sqrt{\left(\mat{Cov}(\vec{p}_{\rm
        MLE})\right)_{ii}} \,.
\end{equation*}

\section{Derivation of the effective degrees of freedom $\tilde K$}
\label{suppsec:effdof}

In the main manuscript the probability distribution for the value of
$\chi^2(\vec{p})$ given $K$ Gaussian process (GP) predictions is approximated by
a non-central chi-squared distribution with $K$ degrees of freedom and a
subsequent transformation to a Gaussian distribution. While $K$ is usually
chosen to be the number of data channels, we define the \emph{effective} number
of degrees of freedom to be a number $\tilde K$ that when used as a parameter of
the probability distribution maximizes the probability density of all $M$
observations $\vec{Y} = [\chi^2(\vec{p}_{1}), \dots, \chi^2(\vec{p}_{M})]\trans$
of the model function $\vec{f}$, i.e.
\begin{equation*}
  \tilde K = K_{\rm MLE} = \underset{K}{\rm arg\,max}\, \mathcal{L}(K; \vec{Y}) \,,
\end{equation*}
where $\mathcal{L}$ is the predictive distribution obtained by approximating the
generalized chi-squared distribution for $\vec{Y}$ twice, as described in
the main manuscript. It is obtained by minimizing the logarithmic likelihood
$\log \, \mathcal{L}$ with respect to the degrees of freedom. Since expressions
for the PDF are involved and its values expensive to
calculate~\cite{mathai1992quadratic,mohsenipour2012distribution}, we opt to find
an approximation and minimize its logarithm instead.

In order to construct the approximate PDF, we first calculate
\begin{equation}
  \label{suppeq:effdof_chi_sq}
  \chi^2_{\rm all} = \sum_{m}^{M}\chi^2(\vec{p}_m) = \sum_{m}^{M} \sum_{k}^{K} \frac{(f_{mk} - t_{k})^2}{\eta_k^{2}} \,,
\end{equation}
where $f_{mk} = f_k(\vec{p}_m)$ is the $k$-th component of the model function
evaluated using the model parameter $\vec{p}_m$, $t_k$ denotes the $k$-th
component of the experimental data, and $\eta_k^2$ denotes the associated
experimental noise variance. The sum can be formulated as a standard scalar
product,
\begin{equation*}
  \chi^2_{\rm all} = \vec{\xi}\trans \vec{\xi} \,,
\end{equation*}
where 
\begin{equation*}
  \vec{\xi} = \left( \frac{f_{11} - t_{1}}{\eta_1}, \dots, \frac{f_{M1} -
    t_{1}}{\eta_1}, \frac{f_{12} - t_{2}}{\eta_2}, \dots, \frac{f_{MK} -
    t_{K}}{\eta_K} \right)\trans
\end{equation*}
is a vector of length $MK$ and follows a multivariate normal distribution,
\begin{equation*}
  \vec{\xi} \sim \mathcal{N}(\vec{\mu}, \mat{M}), \quad \text{with} \quad \vec{\mu} \in \mathbb{R}^{MK} \,, \mat{M} \in \mathbb{R}^{MK\times MK} \,.
\end{equation*}
Using the Kronecker product, the expectation value $\vec{\mu}$ of the vector
can be expressed as
\begin{equation*}
  \vec{\mu} = \left( \frac{\mu_{1} - t_{1}}{\eta_1}, \frac{\mu_{2} - t_{2}}{\eta_2}, \dots, \frac{\mu_{K} - t_{K}}{\eta_K}\right) \otimes \mathbf{1}_{M\times 1} \,.
\end{equation*}
Here, $\mu_{1},\dots,\,u_{K}$ are the mean values of the GPs. The matrix
$\mat{M}$ describes the covariance between the components of the vector
$\vec{\xi}$. Since channels are assumed to be uncorrelated, and only
correlations between observations are considered, $\mat{M}$ is block-diagonal,
\begin{equation*}
  \mat{M} = \diag \left( \left( \frac{\sigma_1^2}{\eta_1^2} \mat{M}_{1} \right), \left( \frac{\sigma_2^2}{\eta_2^2} \mat{M}_{2} \right), \dots, \left( \frac{\sigma_K^2}{\eta_K^2} \mat{M}_{K} \right) \right) \,.
\end{equation*}
Here $\sigma_1^2\dots,\sigma_K^2$ are the variances of the GPs, such that 
$\mat{K}_i = \sigma_i^2 \mat{M}_i$ is the covariance matrix of the $i$-th GP and 
$\mat{M}_i$ are the \emph{unscaled} covariance kernel matrices. 
For the resource reasons discussed in the BTVO
section of the main manuscript, we assume that the $\mat{M}_i$ are shared
between channels, such that $\mat{M}_i = \mat{M}_0$, and therefore
\begin{equation*}
  \mat{M} = \diag \left( \left( \frac{\sigma_1^2}{\eta_1^2} \right), \left( \frac{\sigma_2^2}{\eta_2^2} \right), \dots, \left( \frac{\sigma_K^2}{\eta_K^2} \right) \right) \otimes \mat{M}_{0} \,.
\end{equation*}
Since the matrix $\mat{M}_0$ is unscaled, its variance entries on the diagonal
are equal to $\num{1}$. Too see this, recall from the main manuscript that for a
single Gaussian process we have
\begin{equation*}
(\mat{K}_0)_{ij} = k(\vec{p}_i, \vec{p}_j) \quad \text{and} \quad
k(\mathbf{p},\mathbf{p}') =
\sigma_0^2\left(1+\sqrt{5}r+\frac{5}{3}r^2\right)\exp\left(-\sqrt{5}r\right) \,,
\quad \text{with} \quad r = \sqrt{\sum_{i=1}^N \frac{(p_i-p_i')^2}{l_i^2}} \,.
\end{equation*}
For the diagonal elements of $\mat{K}_0$ we have $\vec{p}_i = \vec{p}_j$, and $r
= 0$, such that $k(\vec{p}, \vec{p}) = \sigma^2_0$, which means that the
diagonal elements of the unscaled covariance matrix are equal to \num{1}.

Analogous to Matsui et al., \cref{suppeq:effdof_chi_sq} is expressed as a linear
combination of squared independent Gaussian random variables
\cite{matsui2019bayesian}. To do this, $\vec{\xi}$ is transformed into a
standard normal,
\begin{equation*}
  \vec{g} = \mat{M}^{-\nicefrac{1}{2}}(\vec{\xi} - \vec{\mu}) \sim \mathcal{N}(\vec{0},\vec{1}_{MK}) \,,
\end{equation*}
and $\mat{M}$ is eigendecomposed as 
\begin{equation*}
  \mat{M} = \mat{P}\trans \mat{Y} \mat{P} \,, 
\end{equation*}
with eigenvectors
\begin{equation*}
  \mat{P} = \vec{1}_{K} \otimes \mat{P}_{\mat{M}_0} \,,
\end{equation*}
where $\mat{P}_{\mat{M}_0}$ are the eigenvectors of the $\mat{M}_0$ matrix, and 
the associated eigenvalues are
\begin{equation*}
  \mat{Y} = \diag (\vec{\lambda}) \quad \text{with} \quad \vec{\lambda} = \left( \frac{\sigma_1^2}{\eta_1^2}, \dots, \frac{\sigma_K^2}{\eta_K^2} \right) \otimes \vec{\lambda}_{\mat{M}_0} \,.
\end{equation*}
Using these definitions
\begin{multline*}
  \chi^2_{\rm all} = \left( \vec{g} + \mat{M}^{-\nicefrac{1}{2}} \vec{\mu}
  \right)\trans \mat{M} \left( \vec{g} + \mat{M}^{-\nicefrac{1}{2}} \vec{\mu}
  \right) = \left( \mat{P}\vec{g} + \mat{P}\mat{M}^{-\nicefrac{1}{2}} \vec{\mu}
  \right)\trans \mat{Y} \left( \mat{P}\vec{g} +
  \mat{P}\mat{M}^{-\nicefrac{1}{2}} \vec{\mu} \right) \\ = \left( \vec{u} +
  \vec{b} \right)\trans \diag (\vec{\lambda}) \left( \vec{u} + \vec{b} \right)
  = \sum_{i}^{MK} \lambda_i (u_i + b_i)^2 = \sum_{i}^{MK} r_i^2 \,,
\end{multline*}
where now $r_i \sim \mathcal{N}(\sqrt{\lambda_i}b_i, \lambda_i)$. To
parameterize the distribution, the eigenvalues of the $\mat{M}$ matrix and the
vector $\vec{b}$ is decomposed as
\begin{equation*}
  \vec{b} = \mat{P} \mat{M}^{-\nicefrac{1}{2}} \vec{\mu} = \left( \frac{\mu_{1}
  - t_{1}}{\sigma_1}, \dots, \frac{\mu_{K} - t_{K}}{\sigma_K}\right) \otimes
  \left( \mat{P}_{\mat{M}_0} \mat{M}_0^{-\nicefrac{1}{2}} \vec{1}_{M \times 1}
  \right) = \vec{b}_1 \otimes \vec{b}_2 \,.
\end{equation*}
Due to the non-unit variances $\lambda_i$, $\chi^2_{\rm all}$ follows a
\emph{generalized} chi-squared distribution. By following the steps taken by
Uhrenholt and Jensen in renormalizing $\chi^2_{\rm all}$~\cite{uhre:2019}, we
can transform the distribution into an approximate \emph{non-central}
chi-squared distribution. The scaling factor for this is the square root of the
mean variance, i.e.
\begin{equation*}
  \gamma 
  = \sqrt{\frac{1}{MK}\sum_{i}^{MK}\lambda_i} 
  = \sqrt{\frac{1}{K}\sum_{k}^{K} \frac{\sigma_k^2}{\eta_k^2} } \sqrt{\frac{1}{M}\sum_{m}^{M} (\vec{\lambda}_{\mat{M}_0})_m}
  = \sqrt{\frac{1}{K}\sum_{k}^{K} \frac{\sigma_k^2}{\eta_k^2} } \,.
\end{equation*}
For the last step, we use that the sum over the eigenvalues of the matrix
$\mat{M}_0$ is equal to its trace, which equates to the number of observations
$M$ since $\mat{M}_0$ is a covariance matrix with unit variances. Each of the
components of $\chi^2_{\rm all}$ is scaled with $\gamma^2$, i.e. $\chi^2_{\rm
all} \to \chi^2_{\rm ren} = \gamma^{-2} \chi^2_{\rm all}$. This then yields
\begin{equation*}
  \gamma^{-2} \chi^2_{\rm all} = \sum_{i}^{MK} s_i^2 \quad \text{with} \quad s_i \sim \mathcal{N}(\sqrt{\gamma^{-2}\lambda_i} b_i, \gamma^{-2} \lambda_i) \,.
\end{equation*}
With $\gamma^{-2} \lambda_i \approx 1$ we therefore assume that $\gamma^{-2}
\chi^2_{\rm all}$ approximately follows a non-central chi-squared distribution,
which is parameterized with $V = MK$ degrees of freedom and a non centrality
parameter
\begin{equation*}
  \kappa = \gamma^{-2} \sum_{i}^{MK} \lambda_i b_i^2 =
  \gamma^{-2}\left(\sum_{i=1}^M (\vec{\lambda}_{\mat{M}_0})_i (\vec{b}_2)_i^2
  \right)\left(\sum_{i=1}^K \frac{\sigma_i^2}{\eta_i^2} \frac{(\mu_i -
  y^\ast_i)^2}{\sigma_i^2}\right) = \gamma^{-2} M \sum_{i}^K \frac{(\mu_{i} -
  t_i)^2}{\eta_i^2} \,.
\end{equation*}
For the last step of the previous equation we use the following identity
\begin{multline*}
\sum_{i=1}^M (\vec{\lambda}_{\mat{M}_0})_i (\vec{b}_2)_i^2 
= \vec{b}_2\trans \diag(\vec{\lambda}_{\mat{M}_0}) \vec{b}_2 
= \mat{1}_{1 \times M} \mat{M}_0^{-\nicefrac{\rm T}{2}} \underbrace{\mat{P}_{\mat{M}_0}\trans \diag(\vec{\lambda}_{\mat{M}_0}) \mat{P}_{\mat{M}_0}}_{\mat{M}_0} \mat{M}_0^{-\nicefrac{1}{2}} \mat{1}_{M \times 1} \\
= \mat{1}_{1 \times M} \underbrace{\mat{M}_0^{-\nicefrac{\rm T}{2}} \mat{M}_0 \mat{M}_0^{-\nicefrac{1}{2}}}_{\mat{1}_{M\times M}} \mat{1}_{M \times 1} = M\,.
\end{multline*}
This distribution is further approximated by means of a normal distribution
where an approximation described by Sankaran \cite{sankaran1959non} is employed,
yielding the approximation of the likelihood function $\mathcal{L}$. In order to
calculate the \emph{effective} degrees of freedom of this distribution, we
minimize the logarithm of the likelihood with respect to the total degrees of
freedom $V$,
\begin{equation*}
  \tilde V = V_{\rm MLE} = \underset{V}{\rm arg\,min}\, \log \left(  \mathcal{L}(V; \gamma^{-2}\chi^2_{\rm all})   \right) \,,
\end{equation*}
where
\begin{equation*}
  \log \left( \mathcal{L}(V; \gamma^{-2} \chi^2_{\rm all})\right) = - \log \left( \varrho \right) - \frac{1}{2} \left( \frac{z - \alpha}{\varrho} \right)^2
\end{equation*}
and
\begin{gather*}
  z = \left( \frac{\gamma^{-2} \chi^2_{\rm all} }{V + \kappa} \right)^{h} \,,
  \qquad \alpha = 1 + h (h - 1) \left( \frac{r_2}{2 r_1^2} -
  (2-h)(1-3h)\frac{r_2^2}{8r_1^4} \right) \,, \qquad \text{and} \\ \varrho = h
  \frac{\sqrt{r_2}}{r_1} \left( 1 - (1-h)(1-3h)\frac{r_2}{4r_1^2} \right) \qquad
  \text{where} \qquad h = 1 - \frac{r_1 r_3}{3 r_2^2} \,.
\end{gather*}
The first three cumulants of the non-central chi-squared distribution
are given as
\begin{gather*}
  r_1(V) = V + \kappa \,, \qquad r_2(V) = 2 ( V + 2 \kappa) \,, \qquad
  \text{and} \qquad r_3(V) = 8 (V + 3 \kappa) \,.
\end{gather*}
A scaling factor $s$ for the actual degrees of freedom for any observation is
then derived by taking $s = V / \tilde V$, such that $\tilde K = K / s$.

We note that the scaling factor $\gamma$ and the non-centrality $\kappa$ do not
depend on entries of the matrix $\mat{M}_0$ of the covariances between the data
points. This means that every observed data point has the same weight for
determining the effective degrees of freedom. As a consequence, the effective
degrees of freedom generally decreases during the minimization of the
chi-squared deviation, since small chi-squared values are more compatible with a
smaller number of degrees of freedom. The smaller values has the advantage that 
it yields a more exploratory behavior once the BTVO algorithm has drawn many 
samples close to an identified local minimum. 
The effective degree of freedom calculated based on a set of uniformly 
distributed samples of the complete parameter space, will generally have a 
larger value that is more representative to the measurement process itself.

\section{MCMC sampling of all parameters of the likelihood function of the MGH17 dataset}

\begin{figure*}[p]
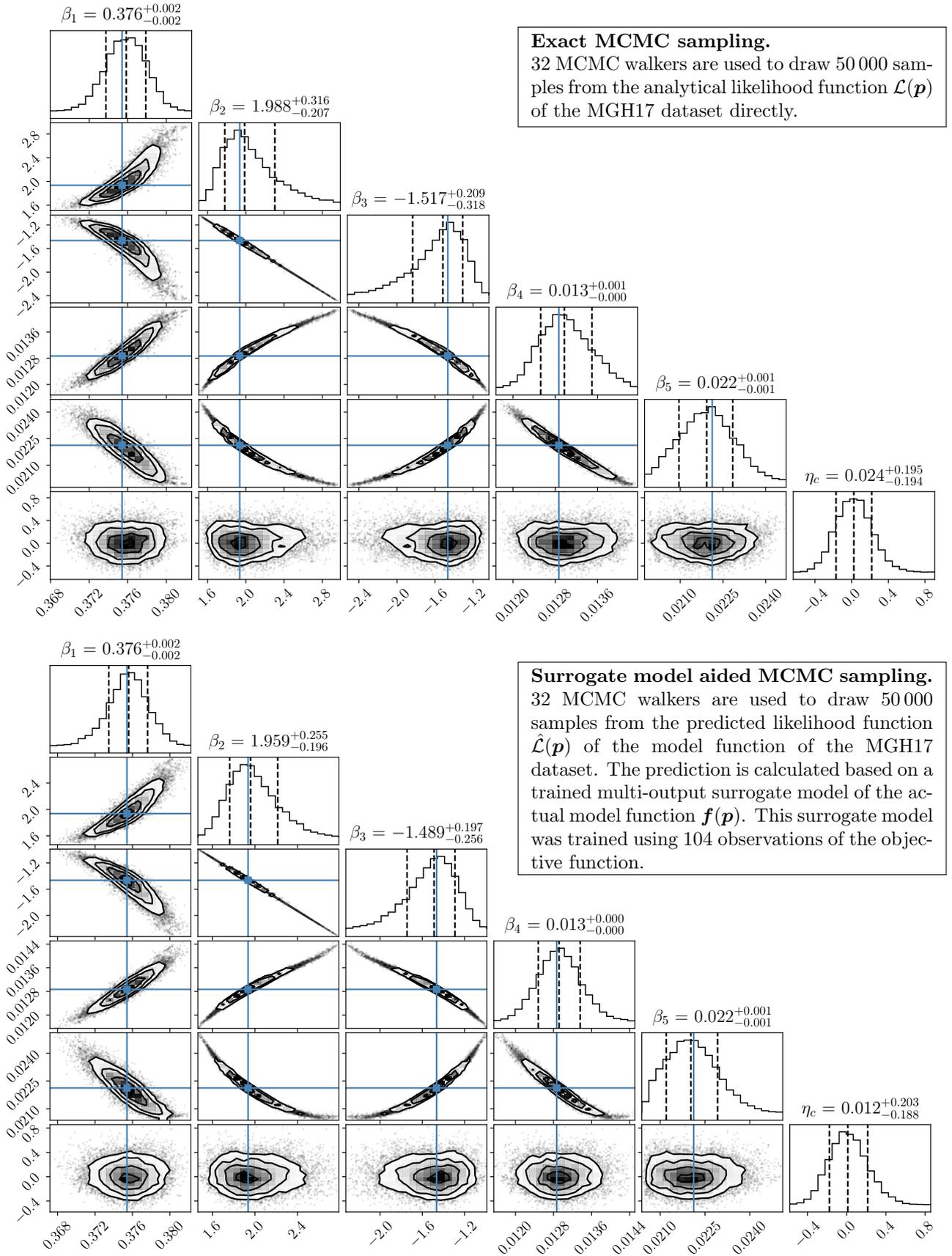

  \centering
  \input{figure_075a6bdb9cd4ec1}
  \input{figure_2280214f480d4e9}
  \caption{The results of MCMC sampling of the likelihood function of the MGH17
    dataset, using the exact likelihood function (left) directly and using a
    trained multi-output surrogate model (right) to calculate a stochastic
    prediction of the likelihood function. A direct comparison of the two
    results reveals that the surrogate model approach is capable of reproducing
    the model parameter distributions as well as revealing correlations between
    the individual model parameters, while using significantly fewer evaluations
    of the model function. The blue lines shown are the certified NIST
    results~\cite{NIST_MGH17}.}
    \label{suppfig:mgh17_mcmc}
\end{figure*}

\cref{suppfig:mgh17_mcmc} shows a comparison of the results of MCMC sampling of
the likelihood functions of the MGH17 dataset, and completes the excerpts shown
in the main manuscript. The top image shows the results of sampling the model
function directly, the bottom image shows the results of sampling a trained
surrogate model of the model function. In addition to the five model parameters
$\beta_1$ to $\beta_5$, an error model scaling parameter $\eta_c$ is shown. A
discussion is found in the main manuscript.

\section{Comparison of GIXRF fit results and experimental values}

\begin{figure}[h]
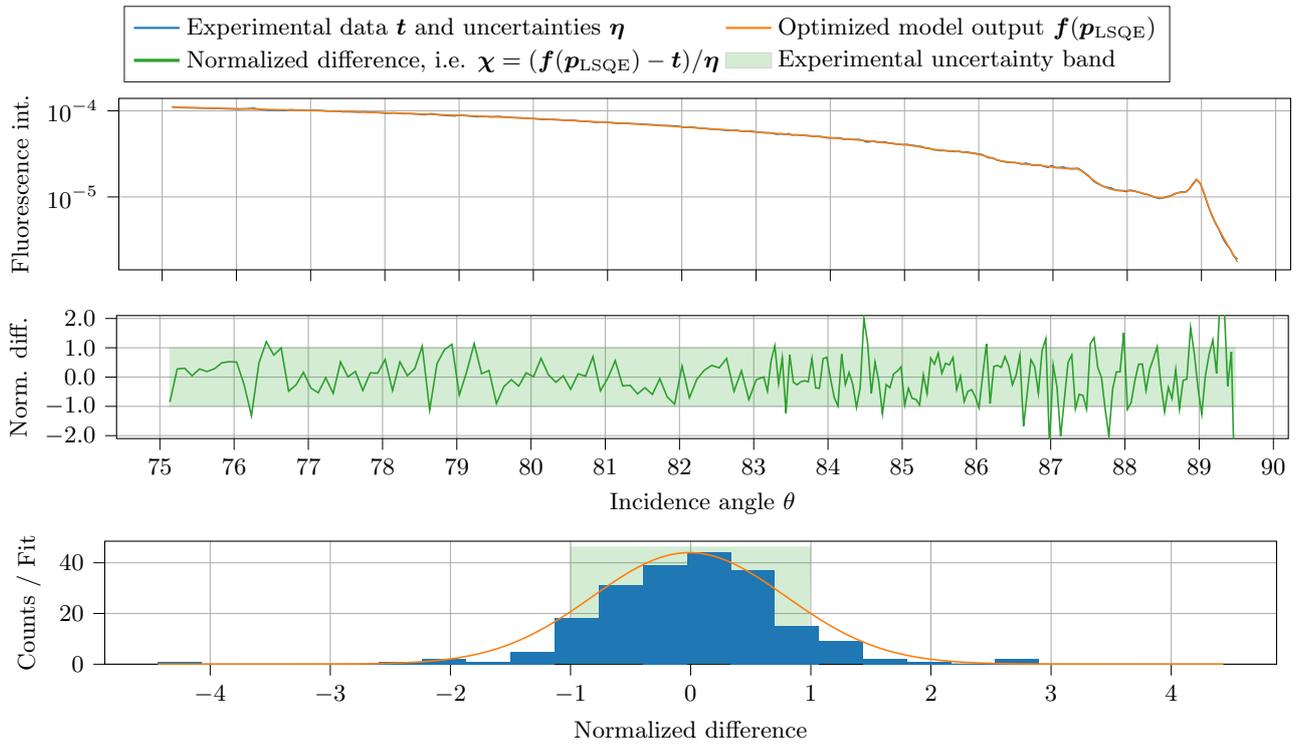

  \centering
  \definecolor{color0}{rgb}{0.12156862745098,0.466666666666667,0.705882352941177}
\definecolor{color1}{rgb}{1,0.498039215686275,0.0549019607843137}
\definecolor{color2}{rgb}{0.172549019607843,0.627450980392157,0.172549019607843}

\begin{tikzpicture}
    \begin{axis}[%
    font=\small,
    hide axis,
    xmin=10,
    xmax=50,
    ymin=0,
    ymax=0.4,
    legend style={draw=white!15!black,legend cell align=left,legend columns=2}
    ]
    \addlegendimage{thick, color0}
    \addlegendentry{Experimental data $\vec{t}$ and uncertainties $\vecc{\eta}$};
    \addlegendimage{thick, color1}
    \addlegendentry{Optimized model output $\vec{f}(\vec{p}_{\rm LSQE})$};    
    \addlegendimage{very thick, draw=color2, fill=color2}
    \addlegendentry{Normalized difference, i.e. $\vecc{\chi}  = ( \vec{f}(\vec{p}_{\rm LSQE}) - \vec{t} ) / \vecc{\eta}$};
    \addlegendimage{area legend, fill=color2, opacity=0.2}
    \addlegendentry{Experimental uncertainty band};
    \end{axis}
\end{tikzpicture}
  \input{figure_796843290a49099}
  \input{figure_c6f974740f5fc73}
  \input{figure_05e6e640666dcfb}
  \caption{
  A comparison of the optimized model results $\vec{f}(\vec{p}_{\rm LSQE})$ and
  the GIXRF experimental data $\vec{t}$. Due to the quality of the fit, the
  experimental data and the optimized model results are indistinguishable. The
  measurement uncertainty $\vec{\eta}$ is approximately two orders of magnitude
  smaller than the experimental data, and is therefore indiscernible in the top
  image. The middle image shows the difference of the model outputs and
  experimental measurements, normalized to the measurement uncertainty. The
  measurement uncertainty is shown as a green bar. The bottom image shows a
  histogram of the relative differences of the middle image, overlaid with a
  (scaled) Gaussian fit of the histogram data.}
  \label{suppfig:gixrf_fit_comparison}
\end{figure}

In \cref{suppfig:gixrf_fit_comparison} the output of the model at the optimized
reconstructed parameter value $\vec{p}_{\rm LSQE}$ is shown in comparison to the
experimental observations. As the model output and the experimental observations
in the top image are indistinguishable, the middle image shows the difference of
the two datasets, relative to the experimental uncertainties. In order to
highlight the quality of the fit, the experimental uncertainties are highlighted
as a green shaded band centered around the zero-difference line. The bottom
image finally shows a histogram of the normalized differences of the middle
image, overlaid with a (scaled) Gaussian fit. As reference, the experimental
uncertainty band is again shown as a shaded green band.

Most of the outputs of the optimized model function are within the uncertainties
of the experimental observations. This is reflected in the Gaussian fit of the
histogram of the normalized differences, which has distribution parameters mean
$\mu = \num{-0.012}$ and standard deviation $\sigma = \num{0.81}$.

\section{Full algorithmic overview}
\label{suppsec:algo_overview}

A full algorithmic overview over the BTVO method and the subsequent MCMC
sampling of the trained surrogate is given in \cref{alg:btvo,alg:mcmc} on
\cpageref{alg:btvo,alg:mcmc}.

\begin{algorithm}[ht]
  \caption{An algorithm overview of the Bayesian target-vector optimiziation scheme.}
  \label{alg:btvo}
  \begin{algorithmic}
    \Procedure{BTVO}{Model function $\vec{f}(\vec{p})$, optimization domain $\mathcal{P}$, target quantities $\vec{t}$ and $\vec{\eta}$}

    \State $X \gets $ Total optimization budget
    \Comment \parbox[t]{.3\linewidth}{E.g. total allowed optimization time, total number of iterations, \dots}

    \State $K \gets \mathrm{dim}(\vec{t})$
    \Comment \parbox[t]{.3\linewidth}{Number of data channels}

    \State $\vec{P} \gets (\mathrm{dim}(\mathcal{P}) + 1) \text{ points from e.g. Sobol sequence in } \mathcal{P}$
    \Comment \parbox[t]{.3\linewidth}{Create initial evaluation candidates.}

    \State $\vec{Y} \gets \vec{f}(\vec{P})$
    \Comment \parbox[t]{.3\linewidth}{Evaluate forwards model for initial candidates}

    \State $M \gets \mathrm{dim}(\mathcal{P}) + 1$
    \Comment \parbox[t]{.3\linewidth}{Evaluations so far.}

    \While{Optimization budget $X \geq \num{0}$}

    \If{$M < M_{\rm Hyp}$ \textbf{and} threshold criterion met}
    \Comment \parbox[t]{.3\linewidth}{For the threshold criterion see
      \cite[Equation (14)]{garcia2018shape}.}

    \State Determine length scales $\vec{l}$ for covariance matrix as MLE
    \Comment \parbox[t]{.3\linewidth}{Determined by averaging across all data channels.}

    \EndIf

    \State Calculate covariance kernel matrix $\mat{K}$

    \State Train $K$ Gaussian processes $\vec{\mathrm{GP}}(\vec{f}; \vec{\mu}, \mat{K})$

    \State Determine effective DoF $\tilde{K}$ from $\vec{Y}$
    \Comment \parbox[t]{.3\linewidth}{See \cref{suppsec:effdof}}

    \State Determine approximate probability distribution $\chi^2(\tilde{K}, \lambda(\vec{p}), \vec{t}, \vec{\eta})$
    \Comment \parbox[t]{.3\linewidth}{For $\lambda(\vec{p})$ see Equation (11) in main manuscript.}

    \State $\vec{p}_{m+1} \gets \arg \max \alpha_{\rm LCB}(\vec{p})$
    \Comment \parbox[t]{.3\linewidth}{Maximize acquisition function using
      approximate probability distribution $\chi^2(\tilde{K},
      \lambda(\vec{p}))$.}

    \State $\vec{y}_{m+1} \gets \vec{f}(\vec{p}_{m+1})$

    \State $\vec{P} \gets \vec{P} \cup \vec{p}_{m+1}$
    \State $\vec{Y} \gets \vec{Y} \cup \vec{y}_{m+1}$

    \State $M \gets M + 1$

    \State $X \gets X - \Delta X$
    \Comment \parbox[t]{.3\linewidth}{Update optimization budget (e.g. reduce remaining iterations, subtract spent time, \dots)}

    \EndWhile

    \State Determine $\vec{p}_{\rm LSQE} = \underset{\vec{p} \in \vec{P}}{\arg\min} \nicefrac{(Y_{i}(\vec{p}) - t_{i})^2}{\eta_i^2}$
    \Comment \parbox[t]{.3\linewidth}{Parameter for which the model value minimizes the chi square value.}

    \State Determine Gaussian uncertainties $\varepsilon_{\vec{p}_{\rm LSQE}}$ of $\vec{p}_{\rm LSQE}$
    \Comment \parbox[t]{.3\linewidth}{C.f. e.g. \cref{suppsec:levenberg-marquardt}.}

    \State \textbf{return} $\vec{p}_{\rm LSQE} \pm \varepsilon_{\vec{p}_{\rm LSQE}}$, $\vec{\mathrm{GP}}$
    \Comment \parbox[t]{.3\linewidth}{Surrogate model is returned for use in \cref{alg:mcmc}.}
    
    \EndProcedure
  \end{algorithmic}
\end{algorithm}

\begin{algorithm}[h]
  \caption{MCMC of expensive model function using Gaussian process surrogates.}
  \label{alg:mcmc}
  \begin{algorithmic}
  
    \Procedure{MCMC}{Model function $\vec{f}(\vec{p})$, surrogate models from optimization step $\vec{\mathrm{GP}}$, target accuracy $\sigma_{\rm min}$, least-square estimate $\vec{p}_{\rm LSQE}$}

    \State $\overline{\sigma}_{\rm max} = \infty$
    
    \While{$\overline{\sigma}_{\rm max} \geq \sigma_{\rm min}$}

      \State Determine covariance matrix $\mat{Cov}(\vec{p}_{\rm LSQE})$ at $\vec{p}_{\rm LSQE}$

      \State Draw $S$ random samples from $\mathcal{N}(\vec{p}_{\rm LSQE}, \mat{Cov}(\vec{p}_{\rm LSQE}))$

      \State Determine $\overline{\sigma}_{\rm max} = \underset{\vec{p} \in \{ \vec{p}_{1}, \dots, \vec{p}_{S} \} }{\max} \frac{1}{K} \sum_{k=1}^{K} \sigma_{k}(\vec{p})$

      \If{$\overline{\sigma}_{\rm max} < \sigma_{\rm min}$}
        \State \textbf{break}
        
      \EndIf

      \State Determine $\vec{p}_{\rm max} = \underset{\vec{p} \in \{ \vec{p}_{1}, \dots, \vec{p}_{S} \} }{\arg\max} \frac{1}{K} \sum_{k=1}^{K} \sigma_{k}(\vec{p})$
      \Comment \parbox[t]{.35\linewidth}{$\sigma_{k}(\vec{p})$ is uncertainty from $\vec{\mathrm{GP}}_{k}$ at position $\vec{p}$.}
    
      \State $\vec{y}_{\rm new} \gets \vec{f}(\vec{p}_{\rm max})$

      \State $\vec{P} \gets \vec{P} \cup \vec{p}_{\rm max}$
      \State $\vec{Y} \gets \vec{Y} \cup \vec{y}_{\rm new}$

      \If{$M < M_{\rm Hyp}$ \textbf{and} threshold criterion met}
      \Comment \parbox[t]{.35\linewidth}{For the threshold criterion see
        \cite[Equation (14)]{garcia2018shape}.}

        \State Determine length scales $\vec{l}$ for covariance matrix as MLE
        \Comment \parbox[t]{.35\linewidth}{Determined by averaging across all data channels.}

      \EndIf

      \State Calculate covariance kernel matrix $\mat{K}$

      \State Train $K$ Gaussian processes $\vec{\mathrm{GP}}(\vec{f}; \vec{\mu}, \mat{K})$

    \EndWhile

    \State Draw $N_{S}$ MCMC samples using $N_{W}$ walkers from the trained surrogate $\vec{\mathrm{GP}}$
    
    \State \textbf{return} Results from processing $N_S$ MCMC samples
    
    \EndProcedure
  \end{algorithmic}
\end{algorithm}

\section{Further benchmarks}
\label{suppsec:further_benchmarks}

In order to expand on the benchmarks shown in the main manuscript, the there
employed methods are used in the reconstruction of the model parameters of
several additional synthetic benchmark problems from the NIST Standard Reference
Database \cite{NIST_StRD}. For the methodology we refer to the main manuscript,
in particular to the section regarding the benchmarks.

Investigated were the problems Benett5 \cite{NIST_Benett5} (three model
parameters, $\num{154}$ model observations), BoxBOD \cite{NIST_BoxBOD} (two
model parameters, six model observations), Eckerle4 \cite{NIST_Eckerle4} (three
model parameters, $\num{35}$ model observations), ENSO \cite{NIST_ENSO} (nine
model parameters, $\num{168}$ model observations), Hahn1 \cite{NIST_Hahn1}
(seven model parameters, $\num{236}$ model observations), and Thurber
\cite{NIST_Thurber} (seven model parameters, $\num{37}$ model observations). The
metric used to measure the performance of the reconstruction algorithms was the
distance of the reconstructed parameter at each iteration to the certified
values $\vec{p}_{\rm Certified}$, in terms of the certified standard deviations,
both as given in the cited sources. We assume no knowledge about possible
multi-modalities of the model functions, and instead treat the as complete
black-boxed models.

In all cases the optimization budget was set to 500 iterations, in the case of
the BoxBOD dataset only the first $\num{250}$ iterations are shown, since the
remainder of the iterations offered no further insights into the convergence
behavior.

\begin{figure}
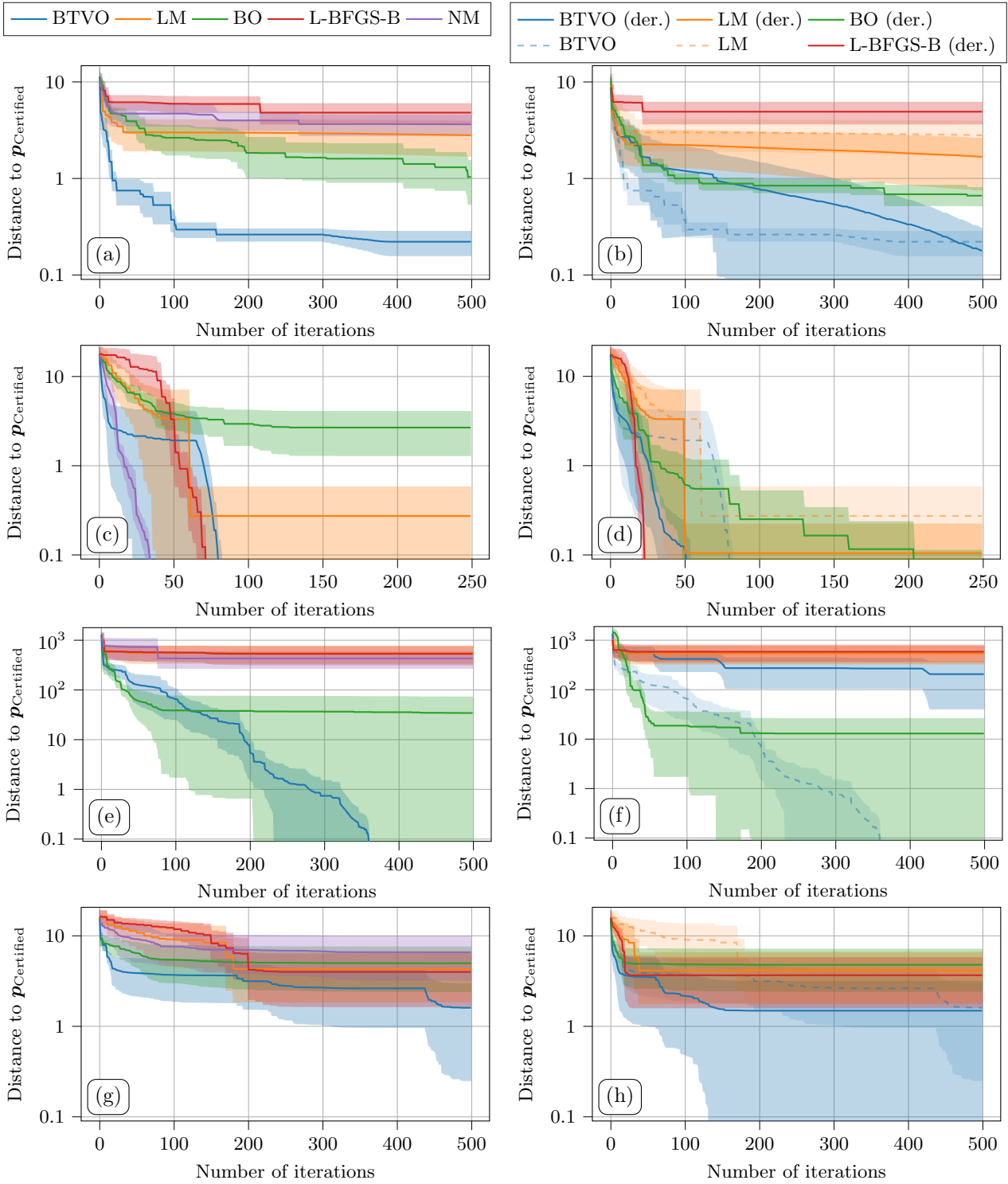

    \centering
    
    \begin{subfigure}[t]{0.49\textwidth}
      \vskip 0pt
      \centering
      \definecolor{color0}{rgb}{0.12156862745098,0.466666666666667,0.705882352941177}
\definecolor{color1}{rgb}{1,0.498039215686275,0.0549019607843137}
\definecolor{color2}{rgb}{0.172549019607843,0.627450980392157,0.172549019607843}
\definecolor{color3}{rgb}{0.83921568627451,0.152941176470588,0.156862745098039}
\definecolor{color4}{rgb}{0.580392156862745,0.403921568627451,0.741176470588235}

\begin{tikzpicture}
    \begin{axis}[%
    font=\small,
    hide axis,
    xmin=10,
    xmax=50,
    ymin=0,
    ymax=0.4,
    legend style={draw=white!15!black,legend cell align=center,legend columns=-1}
    ]
    \addlegendimage{thick, color0}
    \addlegendentry{BTVO};
    \addlegendimage{thick, color1}
    \addlegendentry{LM};
    \addlegendimage{thick, color2}
    \addlegendentry{BO};
    \addlegendimage{thick, color3}
    \addlegendentry{L-BFGS-B};
    \addlegendimage{thick, color4}
    \addlegendentry{NM};
    \end{axis}
\end{tikzpicture}
    \end{subfigure}
    \begin{subfigure}[t]{0.49\textwidth}
      \vskip 0pt
      \centering
      \definecolor{color0}{rgb}{0.12156862745098,0.466666666666667,0.705882352941177}
\definecolor{color1}{rgb}{1,0.498039215686275,0.0549019607843137}
\definecolor{color2}{rgb}{0.172549019607843,0.627450980392157,0.172549019607843}
\definecolor{color3}{rgb}{0.83921568627451,0.152941176470588,0.156862745098039}

\begin{tikzpicture}
    \begin{axis}[%
    font=\small,
    hide axis,
    xmin=10,
    xmax=50,
    ymin=0,
    ymax=0.4,
    legend style={draw=white!15!black,legend cell align=left,legend columns=3}
    ]
    \addlegendimage{thick, color0}
    \addlegendentry{BTVO (der.)};
    \addlegendimage{thick, color1}
    \addlegendentry{LM (der.)};    
    \addlegendimage{thick, color2}
    \addlegendentry{BO (der.)};
    \addlegendimage{thick, color0, dashed, opacity=0.5}
    \addlegendentry{BTVO};
    \addlegendimage{thick, color1, dashed, opacity=0.5}
    \addlegendentry{LM};
    \addlegendimage{thick, color3}
    \addlegendentry{L-BFGS-B (der.)};
  \end{axis}
\end{tikzpicture}
    \end{subfigure}
    
    \begin{subfigure}[t]{0.49\textwidth}
        \vskip 0pt
        \centering
        \input{figure_c777664d566280e}
    \end{subfigure}
    \begin{subfigure}[t]{0.49\textwidth}
        \vskip 0pt
        \centering
        \input{figure_352bf8a7297fbfa}
    \end{subfigure}

    \begin{subfigure}[t]{0.49\textwidth}
      \vskip 0pt
      \centering
      \input{figure_b203dbde2940d31}
    \end{subfigure}
    \begin{subfigure}[t]{0.49\textwidth}
      \vskip 0pt
      \centering
      \input{figure_0a0b1c81aba7317}
    \end{subfigure}

    \begin{subfigure}[t]{0.49\textwidth}
      \vskip 0pt
      \centering
      \input{figure_e5cb02aa5f9ddb7}
    \end{subfigure}
    \begin{subfigure}[t]{0.49\textwidth}
      \vskip 0pt
      \centering
      \input{figure_52d6f118a24a64e}
    \end{subfigure}

    \begin{subfigure}[t]{0.49\textwidth}
      \vskip 0pt
      \centering
      \input{figure_d760f3dac5635e6}
    \end{subfigure}
    \begin{subfigure}[t]{0.49\textwidth}
      \vskip 0pt
      \centering
      \input{figure_671604952a3909b}
    \end{subfigure}

    \caption{Convergence benchmarks for the benchmark problems Benett5 (a,
        b) \cite{NIST_Benett5}, BoxBOD (c, d) \cite{NIST_BoxBOD}, Eckerle4 (e,
        f) \cite{NIST_Eckerle4}, and ENSO (g, h) \cite{NIST_ENSO} from this NIST
        Standard Reference Database \cite{NIST_StRD}. Shown are on the left the
        convergence plots without the use of derivative information (a, c, e,
        g), and on the right the convergence plots with the use of derivative
        information (b, d, f, h). This discussion can be found in
        \cref{suppsec:further_benchmarks}.}
    \label{suppfig:additional_benchmarks_1}
\end{figure}

\paragraph{Benett5} (see \cref{suppfig:additional_benchmarks_1} a and b) has
three model parameters and contains $\num{154}$ model observations. Both with
and without the use of derivative information the BTVO showed the best
reconstruction performance. However, the use of derivative information has lead
to a slower convergence. Despite this, the end result after 500 iterations was
slightly better than the case in which derivative information was not exploited.
It should also be noted that the conventional BO showed good performance when
comparing to the results in the main manuscript. Here, the use of derivative
information enabled the BO method to reconstruct parameters within one standard
deviation of the certified values.

\paragraph{BoxBOD} (see \cref{suppfig:additional_benchmarks_1} c and d) has two
model parameters and only six model observations. For this problem the
conventional optimization methods Nelder-Mead and L-BFGS-B showed very good
reconstruction performance. This may be due to the small loss of information
when calculating the scalar optimization target $\chi^2$ and the low dimensional
parameter space. Of the native least-square methods the BTVO outperformed LM
both when ignoring derivative information and when actively using derivative
information. For the case where derivative information was actively used, the
conventional BO method eventually outperformed LM. It is possible that LM
converged into a local minimum.

\paragraph{Eckerle4} (see \cref{suppfig:additional_benchmarks_1} e and f) has
three model parameters and $\num{35}$ model observations. When neglecting
derivative information the BTVO method performed the strongest. The second best
performance was achieved by the conventional BO method, which however still only
reached a distance of over $\num{10}$ standard deviations from the certified
values. When taking derivative information into account the performance of the
conventional BO method did not change much, the performance of the BTVO however
drastically worsened. This led to the conventional BO method being the
strongest.

\paragraph{ENSO} (see \cref{suppfig:additional_benchmarks_1} g and h) has nine
model parameters and contains $\num{168}$ model observations. Both when
neglecting and when considering derivative information the BTVO showed the best
reconstruction performance. It did however not reconstruct parameters within one
standard deviation of the certified results.

\paragraph{Hahn1} (see \cref{suppfig:additional_benchmarks_2} a and b) has seven
model parameters and contains $\num{236}$ model observations. This dataset posed
a particularly big challenge for all optimization schemes, as none of the
contenders has been able to reach reconstructed parameter values that were
closer than $\num{100}$ standard deviations to the certified values.

\paragraph{Thurber} (see \cref{suppfig:additional_benchmarks_2} c and d) has
seven model parameters and contains $\num{37}$ model observations. The results
for this dataset are comparable to the ones from the ENSO dataset: no method was
capable of reaching values of less than one standard deviation to the certified
values. For this problem the LM method was most successful, both when
considering the model function without and with derivative information.

\paragraph{To summarize,} the BTVO method showed a robust and efficient
performance, especially when no derivative information is used. For two problems
(Benett5, Eckerle4) it could clearly outperform other methods. When it did not
outperform the other methods (BoxBOD, Thurber), it at least showed comparable
capabilities. For three problems (ENSO, Hahn1, Thurber) none of the methods has
been able to robustly reconstruct the certified values to within one standard
deviation.

\begin{figure}
    \centering
    
    \begin{subfigure}[t]{0.49\textwidth}
      \vskip 0pt
      \centering
      \definecolor{color0}{rgb}{0.12156862745098,0.466666666666667,0.705882352941177}
\definecolor{color1}{rgb}{1,0.498039215686275,0.0549019607843137}
\definecolor{color2}{rgb}{0.172549019607843,0.627450980392157,0.172549019607843}
\definecolor{color3}{rgb}{0.83921568627451,0.152941176470588,0.156862745098039}
\definecolor{color4}{rgb}{0.580392156862745,0.403921568627451,0.741176470588235}

\begin{tikzpicture}
    \begin{axis}[%
    font=\small,
    hide axis,
    xmin=10,
    xmax=50,
    ymin=0,
    ymax=0.4,
    legend style={draw=white!15!black,legend cell align=center,legend columns=-1}
    ]
    \addlegendimage{thick, color0}
    \addlegendentry{BTVO};
    \addlegendimage{thick, color1}
    \addlegendentry{LM};
    \addlegendimage{thick, color2}
    \addlegendentry{BO};
    \addlegendimage{thick, color3}
    \addlegendentry{L-BFGS-B};
    \addlegendimage{thick, color4}
    \addlegendentry{NM};
    \end{axis}
\end{tikzpicture}
    \end{subfigure}
    \begin{subfigure}[t]{0.49\textwidth}
      \vskip 0pt
      \centering
      \definecolor{color0}{rgb}{0.12156862745098,0.466666666666667,0.705882352941177}
\definecolor{color1}{rgb}{1,0.498039215686275,0.0549019607843137}
\definecolor{color2}{rgb}{0.172549019607843,0.627450980392157,0.172549019607843}
\definecolor{color3}{rgb}{0.83921568627451,0.152941176470588,0.156862745098039}

\begin{tikzpicture}
    \begin{axis}[%
    font=\small,
    hide axis,
    xmin=10,
    xmax=50,
    ymin=0,
    ymax=0.4,
    legend style={draw=white!15!black,legend cell align=left,legend columns=3}
    ]
    \addlegendimage{thick, color0}
    \addlegendentry{BTVO (der.)};
    \addlegendimage{thick, color1}
    \addlegendentry{LM (der.)};    
    \addlegendimage{thick, color2}
    \addlegendentry{BO (der.)};
    \addlegendimage{thick, color0, dashed, opacity=0.5}
    \addlegendentry{BTVO};
    \addlegendimage{thick, color1, dashed, opacity=0.5}
    \addlegendentry{LM};
    \addlegendimage{thick, color3}
    \addlegendentry{L-BFGS-B (der.)};
  \end{axis}
\end{tikzpicture}
    \end{subfigure}

    \begin{subfigure}[t]{0.49\textwidth}
      \vskip 0pt
      \centering
      \input{figure_6053dbee1c6845a}
    \end{subfigure}
    \begin{subfigure}[t]{0.49\textwidth}
      \vskip 0pt
      \centering
      \input{figure_01cf1051923a1c0}
    \end{subfigure}

    \begin{subfigure}[t]{0.49\textwidth}
      \vskip 0pt
      \centering
      \input{figure_d58469295cc44a6}
    \end{subfigure}
    \begin{subfigure}[t]{0.49\textwidth}
      \vskip 0pt
      \centering
      \input{figure_4621bdb0160e554}
    \end{subfigure}

    \caption{Convergence benchmarks for the benchmark problems Hahn1 (a, b)
      \cite{NIST_Hahn1}, and Thurber (c, d) \cite{NIST_Thurber} from this NIST
      Standard Reference Database \cite{NIST_StRD}. Shown are on the left the
      convergence plots without the use of derivative information (a, c), and on
      the right the convergence plots with the use of derivative information (b,
      d). This discussion can be found in \cref{suppsec:further_benchmarks}.}
    \label{suppfig:additional_benchmarks_2}
\end{figure}

\newpage

\end{document}